\documentclass{article}
\usepackage[utf8]{inputenc}
\usepackage{color}
\usepackage{amsmath,graphics,float,graphicx,amssymb,subcaption,caption}
\usepackage{jcappub}
\def\la{\lower.5ex\hbox{$\; \buildrel < \over \sim \;$}}
\def\ga{\lower.5ex\hbox{$\; \buildrel > \over \sim \;$}}
\title{The effects of the small-scale behaviour of dark matter power spectrum on CMB  spectral distortion}
\author[1,2]{Abir Sarkar,}
\author[1]{Shiv.K.Sethi,}
\author[3]{Subinoy Das}

\affiliation[1]{Raman Research Institute, CV Raman Ave Sadashivnagar,\\  Bengaluru, Karnataka 560080, India}
\affiliation[2]{Indian Institute of Science,CV Raman Ave, Devasandra Layout,\\ Bengaluru, Karnataka 560012, India}
\affiliation[3]{Indian Institute of Astrophysics,100 Feet Rd, Madiwala, 2nd Block,\\ Koramangala, Bengaluru, Karnataka 560034, India}

\emailAdd{abir@rri.res.in}
\emailAdd{sethi@rri.res.in}
\emailAdd{subinoy@iiap.res.in}

\begin{document}

\abstract{After numerous astronomical and experimental searches, the precise particle nature of dark matter is still unknown. The standard Weakly Interacting Massive Particle(WIMP) dark matter, despite successfully explaining the large-scale features of the universe, has long-standing small-scale issues. The spectral distortion in the Cosmic Microwave Background(CMB) caused by Silk damping in the pre-recombination era allows one to access information on a range of small scales $0.3 \, {\rm Mpc} < k < 10^4 \, \rm Mpc^{-1}$, whose dynamics can be precisely described using linear theory. In this paper, we investigate the possibility of using the Silk damping induced CMB spectral distortion as a probe of the small-scale power.  We consider four suggested alternative dark matter candidates---Warm Dark Matter (WDM), Late Forming Dark Matter (LFDM), Ultra Light Axion (ULA) dark matter and Charged Decaying Dark Matter (CHDM);  the matter power in all these models deviate significantly from the $\Lambda$CDM model at small scales.  We compute the spectral distortion of CMB for these alternative models and compare our results with the $\Lambda$CDM model. We show that the main impact of alternative models is to alter the sub-horizon evolution of the Newtonian potential which affects the late-time behaviour of spectral distortion of CMB. The  $y$-parameter diminishes  by a few percent as compared to the $\Lambda$CDM model for a range of parameters of these models: LFDM for  formation redshift $z_f = 10^5$ (7\%);  WDM for  mass $m_{\rm wdm} = 1 \, \rm keV$ (2\%); CHDM for decay redshift $z_{\rm decay} = 10^5$ (5\%); ULA for mass $m_a = 10^{-24} \, \rm eV$ (3\%). \textcolor{black}{This effect from the pre-recombination era can be masked by orders of magnitude higher $y$-distortions generated by late-time sources, e.g. the Epoch of Reionization and tSZ from the cluster of galaxies.} We also briefly discuss the detectability of this deviation in light of the upcoming CMB experiment PIXIE, which might have the sensitivity to detect this signal from the pre-recombination phase.  }

\maketitle

\section{Introduction}\label{intro}

After intensive searches throughout the world over decades, the nature of dark matter is yet to be confirmed. The existence of this component, whose presence is revealed only through its gravitational interaction, is well established by many observations covering a broad range of length scales and epochs of the universe. This list contains cosmic microwave background (CMB) anisotropy experiments \cite{Ade:2015xua,Hinshaw:2012aka,Sievers:2013ica}, large scale structure surveys \cite{Beutler:2016ixs,Tegmark:2006az,Tegmark:2003ud}, the study of the galaxy rotation 
curves \cite{Begeman:1991iy}, cosmological weak gravitational lensing observations \cite{Bartelmann:1999yn,Clowe:2006eq}, etc. 

 One of the leading candidates for dark matter, the Weakly Interacting Massive Particle (WIMP) or the traditional cold dark matter (CDM), is inspired by the well-known WIMP miracle\cite{Craig:2015xla}. The supersymmetric extension of the standard model of particle physics
gives rise to a particle with self-annihilation cross-section $\langle \sigma v\rangle \simeq 3 \times 10^{-26}\rm{cm^3s^{-1}}$ and mass in the range of 100 GeV, which coincidentally produces the correct present abundance of dark matter. Inspired by this discovery, a lot of direct \cite{Angloher:2011uu,Aprile:2010um,Ahmed:2010wy}, indirect \cite{Adriani:2010rc,FermiLAT:2011ab,Barwick:1997ig,Aguilar:2007yf} and collider \cite{Goodman:2010yf,Fox:2011pm} searches have been performed worldwide but none of these experiments have yet succeeded  in providing consistent information about the particle nature of dark matter. In fact, the results from many of these experiments are found to be in conflict \cite{Hooper:2013cwa} with each other.

Besides, there also exist some long-standing astrophysical problems with the  WIMP. One of them is the cusp-core problem
\cite{deBlok:2009sp}, indicated by the discrepancy between increasing dark matter halo profile (cusp) towards the centre of the galaxy from CDM N-body simulations \cite{Navarro:1995iw}, while observationally relatively flat density profiles are found  \cite{Walker:2011zu}. Another issue with WIMP is the missing satellite  problem\cite{Klypin:1999uc,Moore:1999nt};  N-body simulations of structure formation with CDM produce much more satellite halos of a Milky-Way type galaxy than observed. Another issue is the  ``too big to fail'' 
\cite{Garrison-Kimmel:2014vqa, BoylanKolchin:2011de} problem, which underlines the fact, based on N-body simulations,  that a  majority of the most massive subhalos of the Milky Way are too dense to host any of its bright satellites. Some recent works claim that even when the effects of small scale baryonic physics are included,  these issues may still persist  \cite{Pawlowski:2015qta, Onorbe:2015ija, Sawala:2014xka}. All these problems have inspired a drive to go beyond the standard picture of dark matter and consider alternative candidates, which differ from CDM on galactic scales but must reproduce its success on cosmological scales.
We consider four alternative dark matter models in this work: Warm Dark Matter (WDM), Late Forming Dark Matter (LFDM), Ultra Light Axion Dark Matter (ULADM), dark matter produced by the decay of a heavy charged particle in the early universe (CHDM). 
The best studied alternative dark matter model  is the WDM model (e.g. \cite{Dolgov:2000ew, Viel:2005qj}); we consider only thermally produced WDM in this work. This model  has been proposed to 
alleviate   some of the small-scale problems of the $\Lambda$CDM model  \cite{Bode:2000gq}. WDM has been extensively tested against observations 
at small scales: number of satellites based on N-body simulations
\cite{Polisensky:2010rw, Anderhalden:2012qt, Lovell:2011rd}; The cusp-core problem \cite{Maccio:2012qf, Schneider:2011yu};  Lyman-$\alpha$ forest flux power spectrum \cite{Viel:2005qj}; structure formation constraints based on  hydrodynamic simulations \cite{Baur:2015jsy}; the ``too big to fail'' issue \cite{Lovell:2011rd}. These detailed studies prefer WDM masses (or lower limit on the mass) to lie in the
range: $0.1 \, {\rm keV} < m_{\rm wdm} < 4 \, \rm keV$. 
 These constraints do not reveal a unique picture with respect to thermal  WDM being a possible dark matter candidate and underline the need to investigate alternative models. The dark matter produced by the decay of a heavy charged particle in the early universe(CHDM) has also been proposed to address issues with  small-scale power of the 
$\Lambda$CDM model \cite{Sigurdson:2003vy}. LFDM has its origin in extended neutrino physics \cite{Das:2006ht} while the ULA dark matter originates from the string axiverse scenario 
\cite{Arvanitaki:2009fg}. In these cases,  the dark matter starts behaving as CDM after the epoch of Big Bang Nucleosynthesis(BBN) and before the epoch of Matter-Radiation Equality (MRE)  and have similar features (suppression in small scale power followed by damped oscillations) in matter power spectra. \textcolor{black}{For LFDM and CHDM the suppression happens at scales that enter the horizon before the dark matter is formed. For WDM and ULADM models, the scales that suffer suppression are determined by either free-streaming of the particle (WDM) or variable sound speed of the scalar field (ULADM). Both of these scales depend on the mass of the particle/scalar field.} Recent N-body simulations based on  the LFDM and ULA models suggest that these models offer acceptable solutions to the cusp-core issue while being 
consistent with large-scale clustering data  (\cite{Agarwal:2014qca}
\cite{Schive:2014dra, Marsh:2015wka, Marsh:2015xka}). Cosmological 
constraints favour  the following lower limits on the mass of ULA: $10^{-22} \, \rm {eV} < m_a < 10^{-23} \, \rm eV$ \cite{Hui:2016ltb,Sarkar:2015dib}.

All the models we consider deviate from the $\Lambda$CDM model at small scales. One possible probe of this small scale deviation is the spectral distortion of the CMB caused by Silk damping in the pre-recombination era. This damping pumps entropy into the thermal plasma and can distort the CMB spectra if 
the energy injection occurs  after   $z \simeq  10^6$ \cite{1970Ap&SS...7...20S, 1970Ap&SS...9..368S,1975SvA....18..691I,1995A&A...303..323B}. The acoustic waves on scales $10^4 \, {\rm Mpc^{-1}} < k < 0.3 \,  \rm Mpc^{-1}$ get dissipated before the recombination which means that the CMB spectral distortion can be used to constrain the matter power at these scales (for details e.g. \cite{1977NCimR...7..277D,1991ApJ...371...14D,1991A&A...246...49B,Hu:1994bz,Chluba:2012gq,Chluba:2016aln, Khatri:2011aj,Khatri:2012tw,2012JCAP...06..038K,RubinoMartin:2006ug,Chluba:2008aw,Hill:2015tqa,Chluba:2011hw,Dent:2012ne,Chluba:2012we,Clesse:2014pna}). The COBE-FIRAS results gave the current upper bounds on the spectral distortion parameters: 
$|\mu| \lesssim 9 \times 10^{-5}$ and $|y| \lesssim 1.5 \times 10^{-5}$ \cite{Fixsen:1996nj}. Silk damping generally results in spectral distortion amplitudes many orders of magnitude smaller than these limits. \textcolor{black}{The upcoming experiment 
PIXIE \cite{Kogut:2011xw}  will be able to measure $y \simeq {\rm a \, few \,} \times 10^{-9}$ and $\mu \simeq  10^{-8}$, assuming zero foreground contamination. The foregrounds depend on several physical parameters and marginalisation over all of them will degrade the sensitivity\cite{Abitbol:2017vwa}.}

In this work, we study  CMB distortion parameters for alternative dark matter models that differ
from the $\Lambda$CDM model at small scales as possible probes of these models. For these models, we compute the distortion parameters
using the tight-coupling approximation in the pre-recombination era. 

The paper is organised as follows. In Section \ref{particle}, we provide a brief description of all the models considered in this work. In section \ref{sd}, we review the physical picture of the creation and evolution of spectral distortion in the CMB and describe the relation between the dissipation of acoustic wave and CMB spectral distortion. In section \ref{cosmology}, we discuss in detail the physical basis of how a change in the dark matter model impacts Silk damping in the tight-coupling approximation.  Section \ref{result} contains the main results  of this work.  Section \ref{conclusion} is reserved for the conclusion and prospects of this paper.

Throughout this paper, unless specified otherwise, we use the best-fit cosmological parameters given by Planck \cite{Ade:2015xua}: spatially flat universe with  $\Omega_b = 0.049$,   $\Omega_{\rm dm} = 0.254$ (both at the present epoch), and $h= 0.67$.

\section{Models of dark matter} \label{particle}

In this section, we describe the origins of the models considered in this work by motivating their particle physics origins and briefly discuss their cosmological signatures.

\subsection{WDM}
Warm dark matter (WDM) particles with a common mass of $\simeq 1 \, \rm keV$, inspired by particle physics models of sterile neutrinos, have been advocated as a solution to the small-scale anomalies of CDM. A WDM particle is mainly produced non-thermally before the epoch of BBN by active to sterile neutrino oscillation \cite{Shi:1998km} in the early universe. However, there are models of thermal keV mass Sterile neutrinos which were in thermal equilibrium with some hidden sector particle and decouple with correct dark matter relic density \cite{Das:2010ts}. 
Sterile neutrinos in this mass range cannot be detected in standard WIMP searches at least with current experimental capabilities, but in galactic X-ray data, its imprint can be captured if a DM  sterile neutrino decays to a photon and relativistic active neutrino. This photon would have keV energy which in principal can be detected as a galactic X-ray excess. Recent X-ray anomalies from XMM-Newton and Chandra data can be explained by the decay of a 7~\rm{keV} WDM\cite{Boyarsky:2014jta}.\\

\begin{figure}[H]
\centering
\includegraphics[height = 2.5 in, width = 4.5 in]{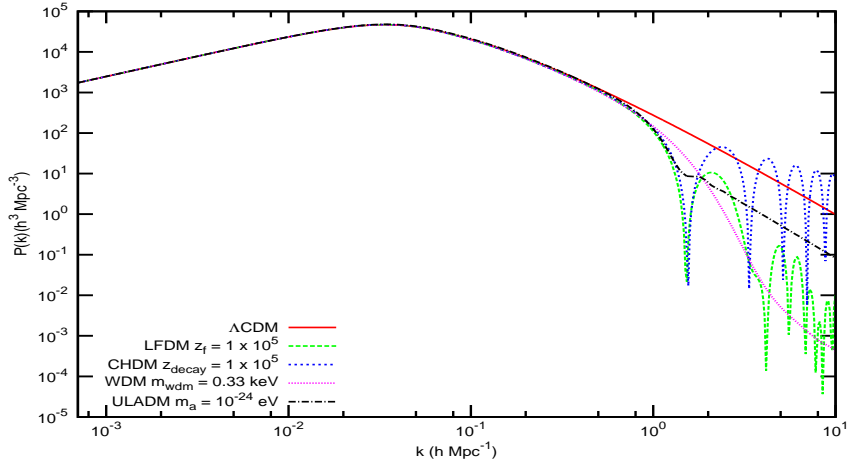}
\caption{Normalized matter power spectra, \textcolor{black}{at z=0}, of the four dark matter candidates considered in this work along with $\Lambda$CDM are shown. For each of the non-standard dark matter candidate, the parameters are chosen such that the power is cut at nearly the same scale $k \simeq 0.3 h \rm {Mpc^{-1}}$. The specifications of the models are mentioned in the legend. }
\label{powerspec}
\end{figure}

In this work, we have considered only thermally produced WDM. In the early universe, WDM is highly relativistic, and its free-streaming scale is the horizon at that time. As time proceeds, it cools down and passing through its semi-relativistic phase it finally becomes non-relativistic CDM. The free-streaming scale during its semi-relativistic phase is given by \cite{Bode:2000gq} 
\begin{equation}
  k_{\rm fs} \simeq \left(\frac{0.3}{\Omega_{\rm wdm}} \right)^{0.15} \,\, \left(\frac{m_{\rm wdm}}{\rm{keV}} \right)^{1.15} \rm {Mpc^{-1}}
\end{equation}
Here $\Omega_{\rm wdm}\equiv \rho_{\rm wdm}/\rho_c$ is the relic density of WDM and $m_{\rm wdm}$ is the mass. In the  matter power spectrum produced by WDM, the power is cut at the WDM free-streaming scale as compared to the standard $\Lambda$CDM model  and the decrement is smooth in nature owing to the fact that the thermal
velocity of  WDM decays as $1/a$ and is not small  enough to be clustered at small scales for a long time  (Figure~\ref{powerspec}). Clearly, lighter WDM will cut power at larger scales (e.g. \cite{Bode:2000gq}). This free-streaming also suppresses the formation of low-mass halos or sub-halos and its finite phase-space density prevents the development of density cusps\cite{Viel:2005qj, Colin:2000dn}. In this work we consider WDM particles in the mass range:  $0.3 \, {\rm keV}<  m_{\rm wdm} < 5 \, {\rm keV}$.

\subsection{LFDM}\label{lfdm}
In this scenario, the dark matter is formed as a consequence of phase transition of a scalar field. The scalar field, which was initially trapped in a metastable minimum by thermal effects or due to its interaction with other scalar fields resulting in a hybrid potential in the early universe, makes a phase transition (which is generally between the redshift of BBN and MRE in this model) to the true minima. After this, the scalar field starts oscillating coherently and its equation of state changes from $w=-1$ to $w=0$, making it clump exactly like the CDM. The dynamics of the phase transition is very similar to the case of the hybrid inflation \cite{Linde:1993cn}, where the scalar field plays the role of the waterfall field and behaves like dark matter after the phase transition. The hybrid potential for two scalar fields in the context of LFDM was derived in \cite{Das:2006ht, Fardon:2005wc} and is given by
\begin{equation}
\label{eq0.1}
V(\phi_1, \phi_2)= (\lambda \phi_1^2 -\mu^2)^2 + 4 \, \lambda^2 \phi_1^2 \phi_2^2 + M^2 \phi_2^2 
\end{equation}
For a high value of $\phi_2 (T)$  in the early universe, the scalar field is trapped in a metastable minimum $\phi_1 =0$ behaving like the cosmological constant. Later, at a critical redshift $z_f$, $\phi_1$ becomes Tachyonic and rolls down to the true minima. After reaching the true minima, $\phi_1$ exhibits coherent oscillations and behaves exactly like the CDM. For details of this model, we refer the reader to \cite{Das:2006ht}.\\

LFDM can also originate from a Fermionic field which has been proposed in \cite{Das:2012kv}. Here, a relativistic Fermionic fluid like the massless neutrino stops free-streaming at a certain redshift $z_f$ due to the onset of a strong fifth force mediated by a scalar field. As the fermions are massive, eventually the fifth force binds all the fermions within a Compton volume of the mediating scalar, forming nuggets. These nuggets are much heavier than the original fermions, and since their formation, they behave exactly like the CDM. \\ 

For both of the scenarios mentioned above, LFDM gets its initial conditions for evolution from the massless neutrino. If LFDM is formed at $z= z_f$, power is suppressed at a scale $ k = k_{\rm lfdm}$ that entered the horizon at that redshift. Unlike WDM, the suppression is sharp as the dark matter is assumed to form via an almost instantaneous phase transition (Figure~\ref{powerspec}). At $ k < k_{\rm lfdm}$, the matter power spectrum carries damped oscillations, a typical feature exhibited by the massless neutrino. This feature is in contrast to WDM where the power falls monotonically without any oscillation. Earlier the dark matter forms, smaller is the scale where the decrement of power occurs. The cosmology of this model is studied in \cite{Sarkar:2014bca} and it is found that dark matter should form deep inside the radiation dominated era and before $z_f \simeq 0.98 \times 10^5$. In the present work, we consider LFDM having formation redshift in the range $5 \times 10^4 < z_f  < 5 \times 10^5$.

\subsection{ULA}\label{ula}
Another very well studied dark matter is the dark matter from ultra light axion (ULA) fields in the context of string axiverse \cite{Arvanitaki:2009fg}. An axion-like particle as a dark matter candidate can be described by \cite{Marsh:2015xka,Marsh:2010wq,Hu:2000ke,Amendola:2005ad} a  two-parameter model, whose action is 
given by:
\begin{equation}
    \label{eq0.2}
     I= \int d^4x \sqrt{g} \left [\frac{1}{2}F^2 g^{\mu \nu} \partial_{\mu}\phi_a \partial_{\nu}\phi_a - \mu^4 (1- \cos{\phi_a})  \right]
\end{equation}
where $\phi_a$ is a dimensionless and periodic scalar field, represented as $\phi_a \rightarrow \phi_a + 2 \pi$. $F$ and $\mu$ are the two parameters
of the model.  For sufficient small value of $\mu$ (which is the case for dark matter), it can be shown that mass of the scalar is given by $m_a = \frac{\mu^2}{F}$. For a cosmologically and astrophysically acceptable dark matter candidate, a reasonable value for the mass is $m_a \simeq 10^{-22}$ eV. All models of particle physics derived from string theory have several periodic scalar fields such as $\phi_a$ and it has been argued that such a low  mass is quite reasonable from particle physics perspective.\\

ULA obtains its initial conditions after spontaneous symmetry breaking in the early universe and behaves like a coherent scalar field. Early when $H$ is high, the friction term dominates and the field is stuck at some random initial value and behaves like the cosmological constant. Later, when $m_a \sim H (z) $ at a certain redshift, the field rolls off and start oscillating coherently around the nearest minima of the periodic potential and starts behaving like CDM.  The adiabatic perturbations in the scalar field have a momentum-dependent and thus mass-dependent effective sound speed. At scales below the effective sound horizon, perturbations are washed out due to free-streaming, and the matter power spectrum features very similar to LFDM (Figure~\ref{powerspec}) are found in the matter power spectrum. The free-streaming scale is given by \cite{Marsh:2015xka}
\begin{equation}
  k_m \simeq \Bigg(\frac{m}{10^{-33}eV}\Bigg)^{1/3} \Bigg(\frac{100\, \rm{km s^{-1}}}{c}\Bigg)h\, \rm{Mpc^{-1}}.
\end{equation}
This means that lighter axions will push the scale of suppression to a higher value. The cosmologically relevant mass  of ULA  ranges from $10^{-18}$ and can be as tiny as $10^{-33}$ eV.  Some  recent works have put a lower limit on  the 
mass of   ULA: $m_a \lesssim 2.6 \times 10^{-23}$ \cite{Hlozek:2014lca,Sarkar:2015dib}.  In this paper, we consider the mass  range:  $10^{-21} \, {\rm eV}  > m_a > 10^{-25} \, \rm eV$.

\subsection{Charged decaying dark matter} \label{sec:chdm}
This model, its variants, and  their cosmological implications have been  investigated in detail  \cite{Sigurdson:2003vy,Kaplinghat:2005sy,Profumo:2004qt,Kohri:2009mi,Kamada:2016qjo,2015PhRvD..91b3512F}.  We consider a model  in which a  heavy  negatively charged particle of mass $M_{\rm ch}$
decays into a heavy neutral particle of mass $M_{\rm neu}$  and a relativistic electron (supersymmetric models in which a selectron decays into an electron and a gravitino might achieve this scenario \cite{Sigurdson:2003vy}). These two masses and the decay time $\tau$ parameterize the model. The decay time  and the mass difference between the two heavy particles $\Delta M =M_{\rm ch} -M_{\rm neu}$ are  tightly constrained  because the relativistic electron thermalizes with electron-photon coupled system, thereby causing spectral distortion of CMB if the decay time corresponds to redshifts $z_{\rm decay} < 10^6$.  The three-momentum of the relativistic electron $p = \Delta M$ and in the limit all this relativistic energy is transferred to the photon gas, we get the fractional energy increase: 
\begin{equation}
\delta\rho_\gamma/\rho_\gamma \simeq 4.2 \times 10^{-2} \left ({\Omega_{\rm dm}h^2\over 0.11}\right ) \left({10^5 \over 1+z}\right) \left({\Delta M \over M_{\rm ch}}\right)
\label{eq:chdistor}
\end{equation}
 Using the current bounds on $\mu$ and $y$-parameters, we get \textcolor{black}{$\Delta M/ M_{\rm ch} \la 10^{-2}\hbox{--}10^{-3}$} for decay times in the redshift range  $10^6 > z_{\rm decay} >  10^5$. 

This constrains the energy density of the relativistic electrons to be dynamically unimportant and allows us to assume that the masses of the two heavy particles are the same. This means that the main difference between models such as the  LFDM model and the decay charged particle model is that whereas the initial conditions (density and velocity perturbations)  in the former case arise from massless neutrino, they are inherited from the baryon-photon fluid in the latter case. \footnote{This assumption needs further explanation. For a massive charged particle to share the bulk velocity of the baryon-photon fluid, it must be tightly coupled to this fluid which means the time scale of energy exchange between this particle and the photon-baryons fluid should be far smaller than the expansion rate. This time scale $\tau_{\rm chelec} \simeq 100 \,  (10^6 {\rm K}/T)^{3/2} (100 {\rm Gev}/M_{\rm ch}) \, {\rm sec}$ which is much shorter than the expansion rate at redshifts of interest. Another complication, in this case, arises from the possibility that the charged dark matter particle can form an atom with a proton. This hydrogenic atom 
would have binding energy on the order of 24~keV ($\simeq (m_p/m_e) 13.6 \, \rm eV$). These atoms, in turn, could be converted into an ion containing the charged
dark matter particle and doubly ionized helium through charge exchange reactions(for details e.g. \cite{2011piim.book.....D}). Only a tiny fraction of these atoms will form until the epoch there 
are not enough number of energetic 
CMB photons on the tail of black body spectrum to ionize the atom; this epoch corresponds roughly to a temperature $\simeq 200 \, \rm eV$. This could be  an additional source of spectral distortion which we neglect in this paper.} This results 
in a qualitative difference between the two cases as seen in Figure~\ref{powerspec}. While density and velocity perturbations of massless neutrinos decay 
after horizon entry, these perturbations oscillate with nearly constant amplitude for the photon-baryon fluid after horizon entry for $\eta < \eta_{\rm decay}$.  

It is seen  in Figure~\ref{powerspec} that the matter power spectrum
in this case oscillates for scales that are sub-horizon during the pre-decay phase but its value can exceed the matter power spectrum for the $\Lambda$CDM 
model for the same cosmological parameters. 

This can be understood as follows. In the $\Lambda$CDM model, the density 
perturbations of the CDM component, in Synchronous gauge,  $\delta_{\rm cdm} = -h/2$ or they are completely determined by metric perturbations and are independent of velocity perturbations which are zero at all times,  $\theta_{\rm cdm} = 0$. In this case,  the CDM  density perturbations at sub-horizon scales grow logarithmically during the radiation-dominated era. However, in the decaying charged particle model, the initial velocity perturbations of the CDM component (the post-decay neutral particle) are derived from the photon-baryon fluid and constitute an additional source of density perturbations. For wavenumbers at which  the velocity perturbations of the initial conditions combine in phase with density perturbations of the CDM component, the density perturbations can overshoot
perturbations in the $\Lambda$CDM model. On the other hand, this effect also 
serves to increase the decrement as compared to the $\Lambda$CDM model for  wavenumbers at which velocity perturbations act to suppress the growth of density perturbations. 

All the models we consider here leave the matter-radiation equality unchanged and their matter energy densities  at the present epoch are normalized   to $\Omega_{\rm dm} = 0.254$.   In Figure~\ref{powerspec}, the power spectra of all the four models are displayed  and compared to that of $\Lambda$CDM. In the Figure,  we have plotted  models which induce cut in power at  nearly the same  scale $k \simeq 0.3 h\, \rm {Mpc^{-1}}$. The power spectra shown in the figure  have been computed 
 using the  publicly-available codes   \texttt{CMBFAST}  (WDM, CHDM, and LFDM by  modifying this code) and \texttt{axionCAMB} (ULA).
 
 \subsection{\textcolor{black}{Evolution of the transfer functions of different dark matter candidates}}

\begin{figure}
\begin{subfigure}{.5\textwidth}
  \centering
  \includegraphics[width=1.0 \linewidth ]{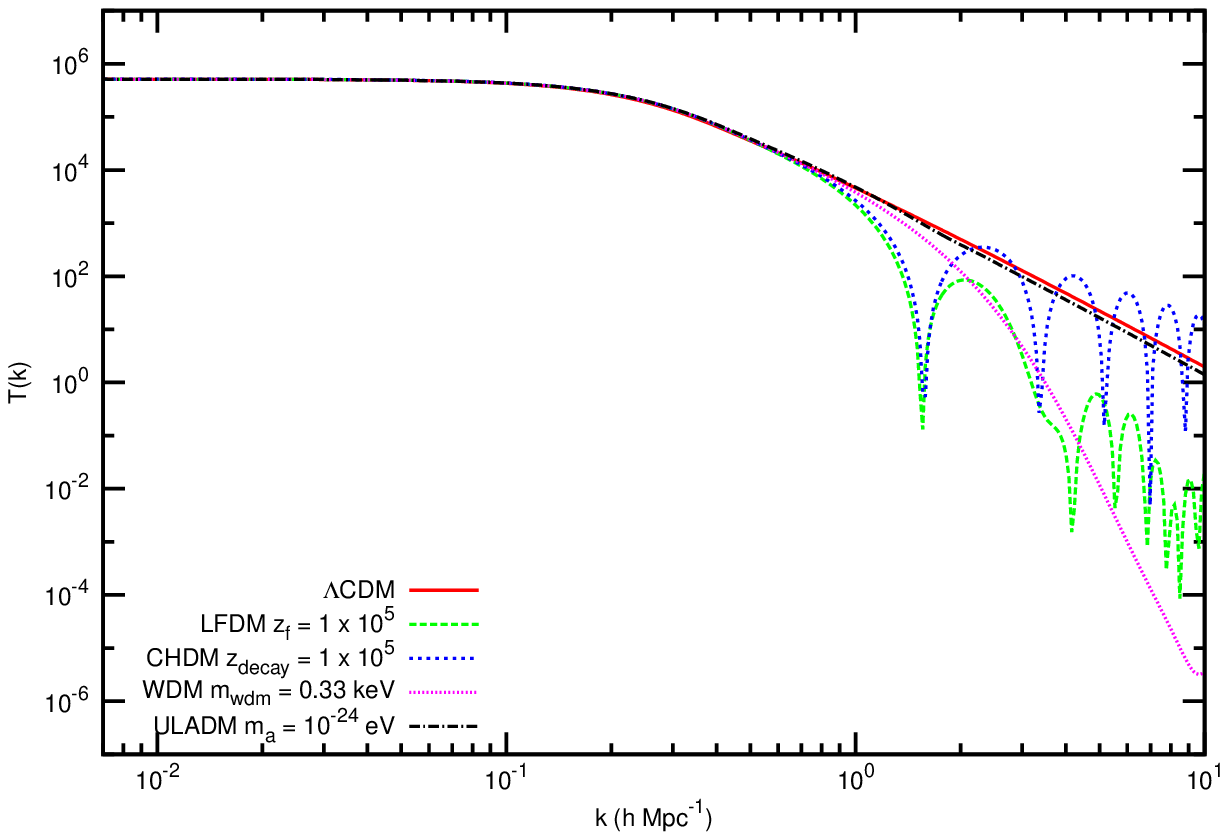}
  \label{fig:tra}
\end{subfigure}%
\begin{subfigure}{.5\textwidth}
  \centering
  \includegraphics[width=1.0 \linewidth ]{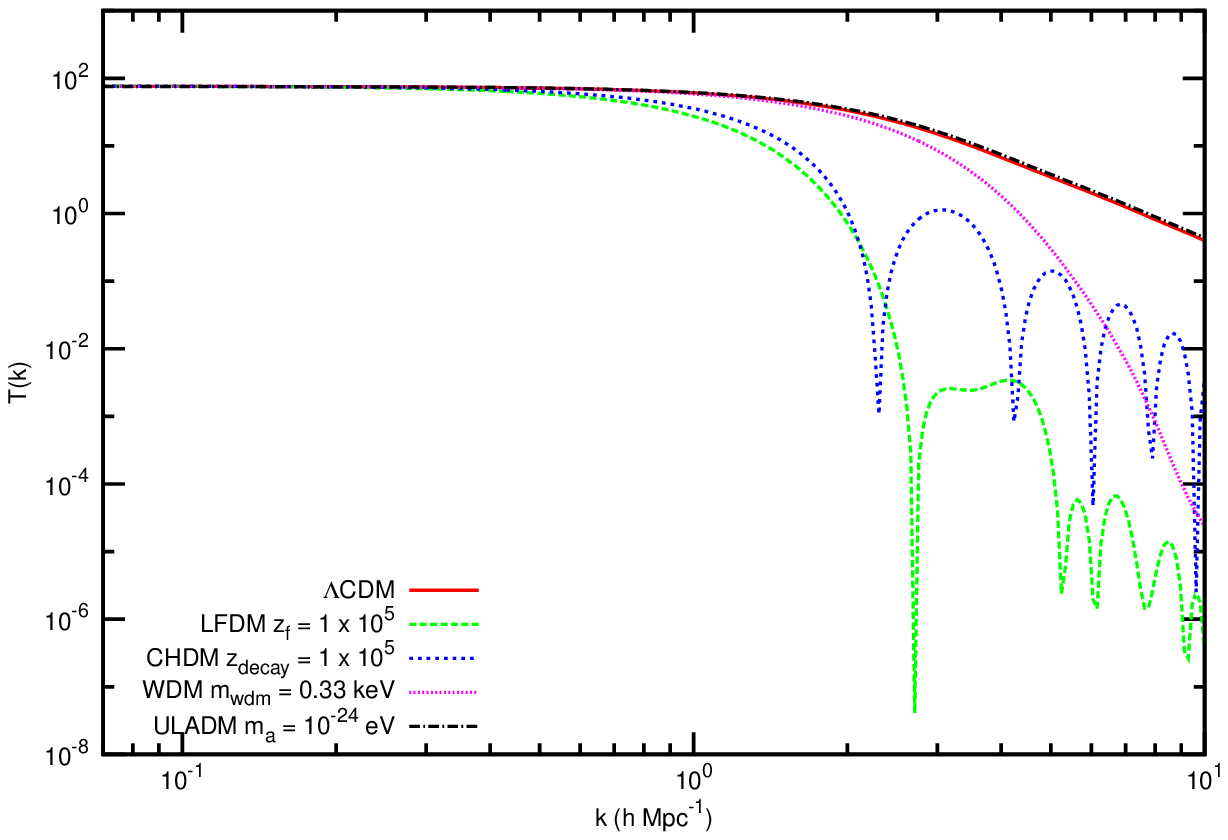}
  \label{fig:trb}
\end{subfigure} \\
\begin{subfigure}{.5\textwidth}
  \centering
  \includegraphics[width=1.0 \linewidth ]{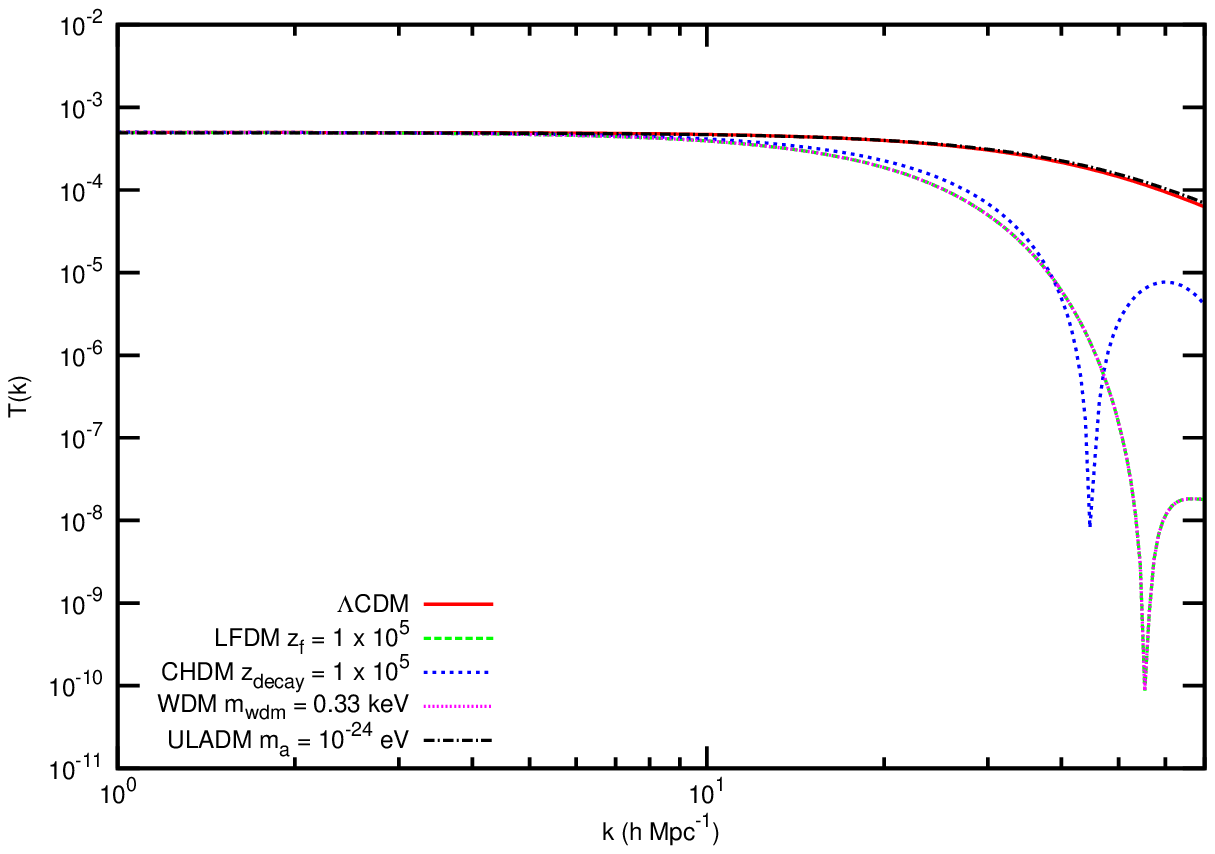}
  \label{fig:trd}
\end{subfigure}%
\begin{subfigure}{.5\textwidth}
  \centering
  \includegraphics[width=1.0 \linewidth ]{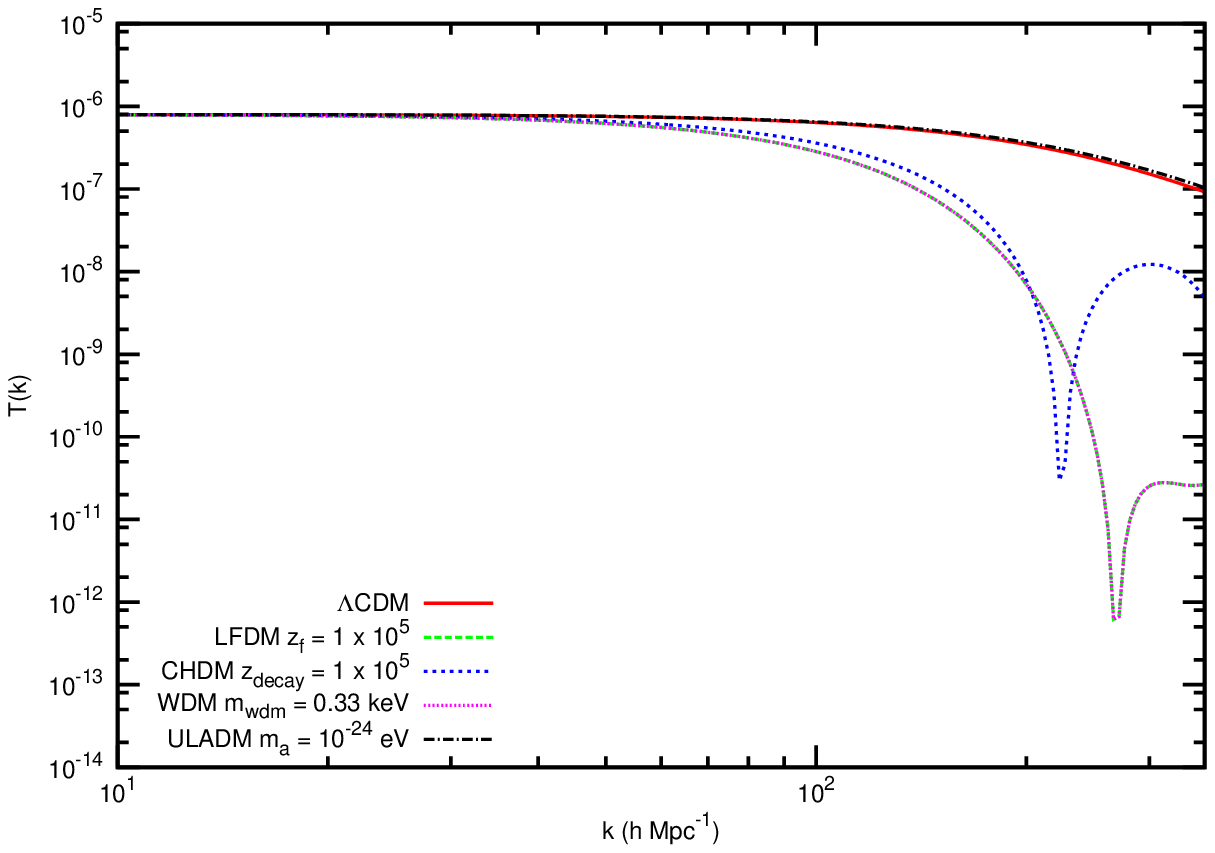}
  \label{fig:tre}
\end{subfigure} 
\caption{The evolution of transfer functions of four dark matter candidates considered in this work along with $\Lambda$CDM. Clockwise from top-left, transfer functions are plotted at $z = 10^4, \, 10^5, \, 2 \times 10^6\, {\rm and} \, 10^7$. For each of the non-standard dark matter candidate, the parameters are chosen such that the power is cut at nearly the same scale $k \simeq 0.3 h \rm {Mpc^{-1}}$. The specifications of the models are mentioned in the legends. }
\label{transfer}
\end{figure}

   \textcolor{black}{In this subsection, we briefly describe how the dark matter transfer function evolves for different dark matter candidates considered in this work. The transfer functions are given in Figure~\ref{transfer} at $z = 10^4, \, 10^5, \, 2 \times 10^6\, {\rm and} \, 10^7$. The model specifications are identical to that used in Figure~\ref{powerspec}. This allows us to motivate the 
discussion in  section~\ref{cosmology}.}  \\

      \textcolor{black}{\textbf{WDM}: In this work, we have considered keV-mass thermally produced WDM candidates, which become non-relativistic at $z \sim 10^6$--$10^7$. So in the very early universe, when they are still relativistic, the transfer function at sub-horizon scales will be similar to that of a relativistic particle like the massless neutrino. As time progresses they become non-relativistic but still cause suppression as compared to the $\Lambda$CDM model at small scales owing to their free-streaming velocity that decays at a slower pace, as $1/a$.  We have shown the evolution of transfer functions for WDM with of mass 0.33~keV in Figure~\ref{transfer}, which is relativistic until $z \simeq 1.4 \times 10^6$, so the sub-horizon features at $z = 10^7$ and 
 $z = 2 \times 10^6$ are essentially the same. WDM with higher masses become non-relativistic earlier, before the onset of the $\mu$-distortion era (discussed in the next section)}.\\

   \textcolor{black}{\textbf{ULADM}: The ULADM forms due to spontaneous symmetry breaking in the early universe and behaves like a coherent scalar field.  Once the mass of the field drops below the Hubble constant, the field starts oscillating coherently around the true minima of the potential and behaves like cold dark matter. We have plotted the evolution of transfer function for $m_a = 10^{-24} {\rm \, eV}$.  ULADM with this mass decouples from the Hubble drag at $z \simeq 2 \times 10^4$ and the suppression occurs  at scales smaller than $k \simeq 0.3 {\rm \, hMpc^{-1}}$. So both at the onset of $\mu$-distortion (Figure~\ref{fig1} in section~\ref{sd}) and at earlier times the sub-horizon transfer functions are the same, as seen from the Figure~\ref{transfer}.}\\
    
   \textcolor{black}{\textbf{LFDM}: As mentioned in subsection~\ref{lfdm}, LFDM is formed due to phase transition in the neutrino sector and gets its initial conditions  from the massless neutrino. In Figure~\ref{transfer}, we show
the evolution of the transfer function for $z_f = 10^5$. We notice that the transfer functions of LFDM and the WDM with mass 0.33~keV are identical at $z = 10^6$ and $10^7$ because at high redshifts both behave like massless neutrinos. 
At $z = 10^5$, the WDM has already become non-relativistic but LFDM continues to be relativistic which explains greater suppression in the LFDM power spectrum. After $z = z_f$, the LFDM behaves as CDM, but WDM continues to cause suppression
of power at small scales owing to its slowly decaying but still significant free-streaming velocities.  At more recent time, i.e., $z = 10^4$, the LFDM transfer function is identical to that of CDM at scales $k \gtrsim 0.3 {\rm hMpc^{-1}}$, the scale that enters the horizon at the formation redshift, but smaller scales still carry the massless neutrino-like features.}\\
    
   \textcolor{black}{\textbf{CHDM}: This model is qualitatively different from the other models as the dark matter candidate, in this case, is charged and coupled to the photon-baryon plasma at early times. The time evolution of CHDM transfer function  is identical to that of baryons at high redshifts with $R=3(\rho_b+\rho_{\rm dm})/(4\rho_\gamma)$ as explained in subsection~\ref{sec:chdm}. At lower redshifts and at scales larger than the scale that entered the horizon at $z_{\rm decay} = 10^5$, its transfer function is identical to CDM. At smaller scales, the transfer function can exceed the matter power  of the $\Lambda$CDM model, as seen in Figure~\ref{transfer} and explained in subsection~\ref{sec:chdm}}.
    
\section{Physics of Spectral Distortion} \label{sd}
\begin{figure}
    \centering
    \includegraphics[height = 4in, width = 6.5in]{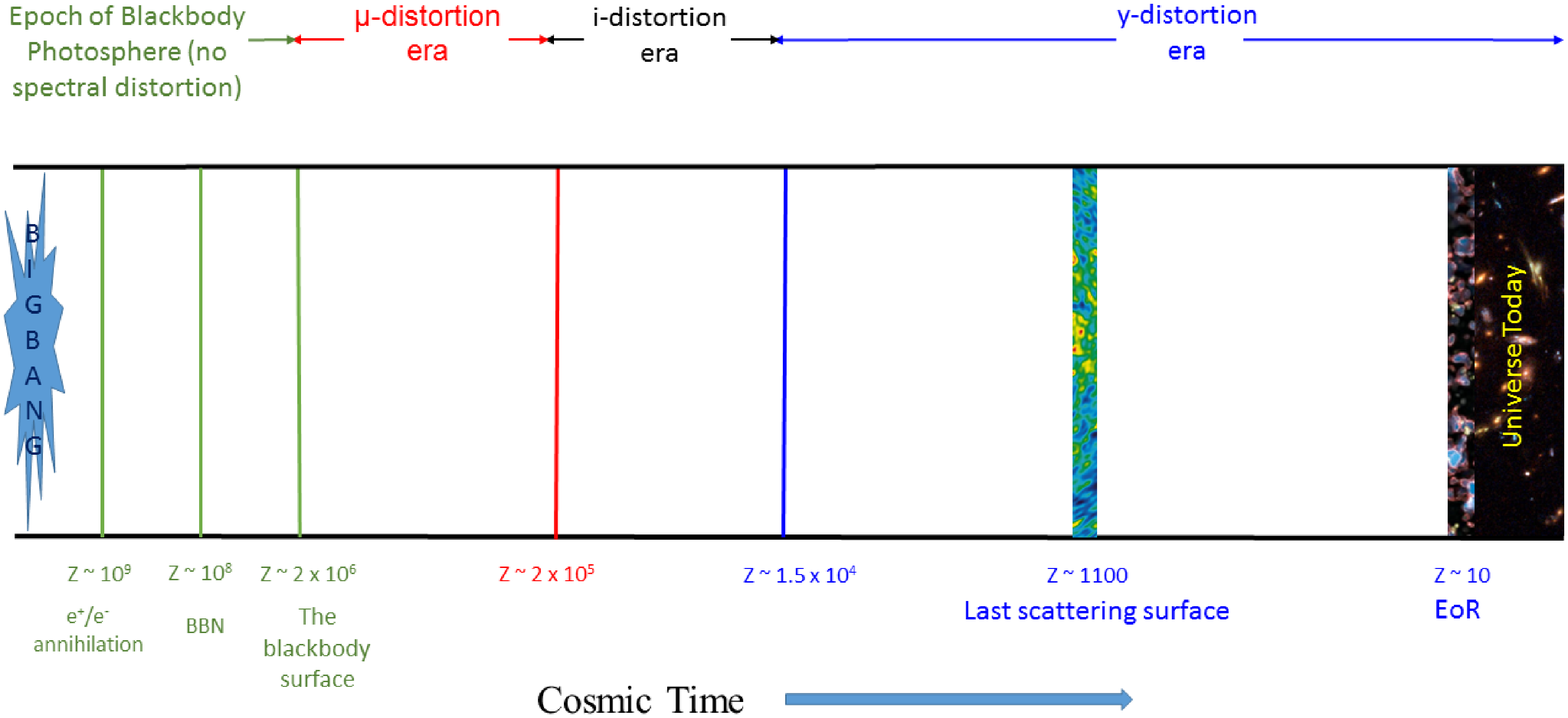}
    \caption{Important epochs of the evolution of spectral distortion in the Cosmic Microwave Background.}
    \label{fig1}
\end{figure}

In this section, the physics of spectral distortion in the CMB will be discussed in more detail. In Figure~\ref{fig1} different epochs related to the evolution of CMB  distortion are shown. There are four phases in this regard, and we provide a brief physical picture of each of these. 

In the early universe, there are two classes of physical processes that can cause energy exchange between electrons and photons and therefore act to equilibrate photon distribution function to a black body. One is photon conserving process, dominated by inverse Compton scattering and photon producing processes which are mediated by double Compton scattering and free-free emission with double Compton scattering dominating the rate of production of photons in the early universe. If both these processes act on time scales shorter than the expansion time scale of the universe, any injection of energy in the universe is rapidly shared between the photons and charged particles, and the CMB spectrum relaxes to a black body  ( see e.g. \cite{1975SvA....18..691I, 1979rpa..book.....R, 1981ApJ...244..392L, Hu:1992dc}). 

This condition is obtained in the first era $z \gtrsim 2 \times 10^6$. Any injection of energy during this era, e.g.  $e^+/e^-$ annihilation and BBN, leaves no trace on the CMB spectrum. 

For $z \lesssim 2 \times 10^6$ the rate of photon production is not large enough to equilibriate the photon distribution function to a black body except at low photon frequencies. This redshift thus approximately marks the time after which the signature of energy injection can no longer be erased. As Compton scattering remains 
an efficient process for energy exchange between photons and electrons, the 
resultant distribution function relaxes to a Bose-Einstein distribution 
with $T_\gamma = T_e$ but with a non-zero chemical potential, $\mu$. This 
arises because Compton scattering is a photon conserving process that cannot 
cause a transition between two Black bodies at two different temperatures.
During this phase, the evolution of the photon distribution function, $n(\nu,t)$ with Compton scattering being the only energy-exchange process, is given 
by the well-known Kompaneets equation:
\begin{equation}
\label{eq1.0}
\frac{\partial n}{\partial t} =\frac{n_e\sigma_Tk_BT_e}{m_ec}
\frac{1}{x_e^2}\frac{\partial}{\partial x_e} x_e^4 \left[\frac{\partial n}{\partial x_e}+(n+n^2)\right ] 
\end{equation}
Here  $T_e$ is the electron temperature, $n_e$ is the number density of electrons, $m_e$ is the mass of the electron and  $x_e = h\nu/k_BT_e$, $\nu$ being the frequency of the photon, is the dimensionless frequency. 
The general equilibrium solution of this equation is given by \cite{1990PhRvD..41..354B, 1970Ap&SS...7...20S}
\begin{equation}
\label{eq1.2}
n(\nu) = \frac{1}{e^{x_e+\mu}-1}
\end{equation}
The chemical potential $\mu$ can be related to the fractional energy  absorbed by the photon gas: $\mu = 1.4 {\Delta\rho_{\gamma}/\rho_{\gamma}}$ (e.g. \cite{1975SvA....18..691I,1970Ap&SS...7...20S}). 

For  $z \la  2 \times 10^5$, the Comptonization process is not efficient enough to bring the photon distribution function into equilibrium, and the equilibrium solution of Kompaneets equation (Eq.~\ref{eq1.2}) is not valid anymore. This redshift marks the end of $\mu$-distortion era. This era is followed  by ``$i$-distortion" and ``$y$-distortion" eras.  We first discuss the physics of $y$-distortion. 

The distribution function   in Eq.~\ref{eq1.2} satisfies the following identity : $\frac{\partial n}{\partial x} = -(n+n^2)$ where $x=h\nu/kT$.  Assuming small  departure from equilibrium, Eq.~\ref{eq1.0} can be written as follows:
\begin{equation}
\label{eq1.5}
\frac{\partial n}{\partial t} = \frac{n_e\sigma_Tk_B(T_e-T)}{m_ec}\frac{1}{x^2}\frac{\partial}{\partial x} x^4 \left[\frac{\partial n}{\partial x}\right ] 
\end{equation}
The time variable of Eq.~\ref{eq1.5} can be modified as a new parameter which denotes the temperature difference between electrons and photons:
\begin{equation}https://www.sharelatex.com/project/5666c560182bc84705fe62a3
\label{eq1.4}
dy = \frac{n_e\sigma_Tk_B(T_e-T)}{m_ec}dt
\end{equation}
Eq.~\ref{eq1.5} can be solved to give \cite{1990PhRvD..41..354B,1969Ap&SS...4..301Z}:
\begin{equation}
\label{eq1.8}
n(x,y) = n(x,0)+\frac{xye^x}{(e^x-1)^2}\left[\frac{x}{\tanh(x/2)}-4\right]
\end{equation}
In this case  the fractional change in photon energy can be related to the 
$y$-parameter as: $y = 1/4 {\Delta\rho_{\gamma}/\rho_{\gamma}}$ \cite{1969Ap&SS...4..301Z}.

\textcolor{black}{Between $\mu$- and $y$-distortion era there exists another era called the ``intermediate ($i$-type) distortion era". In this time, the Comptonization time scale is not short enough to relax the spectrum to equilibrium. Instead, the system settles into a state where the distortion is given by the sum of $\mu$-,$y$- and some residual distortion  \cite{Chluba:2011hw, Chluba:2013vsa}. The thermalization in this era is approximated as a weighted combination of pure $\mu$- and $y$-distortion\cite{Chluba:2013vsa}. The residual distortion ($r$-type) is between  10-30$\%$ of the total distortion. This $r$-type distortion depends sensitively on the time of energy injection, which is not the case for pure $\mu$- or $y$-distortion. A different approach was taken in \cite{Khatri:2012tw} to quantitatively describe the $i$-distortion. The $i$-distortion can be characterised by  a  modified $y$-parameter defined as:
\begin{equation}
\label{eq1.9}
y_{\gamma}(z_{inj},z_{max}) = -\int_{z_{max}}^{z_{inj}} \frac{n_e\sigma_Tk_BT}{m_ecH(1+z)}dz
\end{equation}
The Kompaneets equation is written in terms of this new time parameter and then expanded about $y_{\gamma}$. The solution shows that the distorted spectrum and thus the distortion is dependent on $y_{\gamma}$. According to Eq.~\ref{eq1.9}, $y_{\gamma}$ is sensitive to the time when the energy was injected into the system through $z_{inj}$. This makes the  $i$-distortion able to estimate not only the amount of energy injected but also the time of injection, something that can't be estimated by observing the $\mu$ or $y$-type distortion.}

After $z \sim 1.5 \times 10^4$, $y_{\gamma}$ becomes very small and $y$-distortion epoch commences  and lasts up to present time. Late time phenomena of the universe like reionization \cite{Hu:1993tc,Khatri:2012tw}, heating of photons by warm-hot intergalactic medium \cite{Nath:2001yd,Komatsu:2002wc,Sehgal:2009xv} and SZ effects from groups and clusters of galaxies (\cite{Hill:2015tqa,1991A&A...246...49B} and references therein) also contribute to the $y$-distortion. As long as the amount of distortion is small, the three different  distortions \textcolor{black}{($\mu$+$y$+$r$)} can be linearly added to give the final spectrum.

Various processes like decay of massive particles\cite{Chluba:2011hw,Chluba:2013wsa,Chluba:2013pya}, annihilation of particles \cite{Chluba:2016aln,2005PhRvD..72b3508P,2010MNRAS.402.1195C}, dissipation of acoustic wave \cite{Chluba:2016aln,1970Ap&SS...9..368S,1991ApJ...371...14D,Hu:1994bz,Chluba:2012gq,Pajer:2013oca}, adiabatic cooling of electrons \cite{Chluba:2016aln, Khatri:2011aj} and cosmological recombination radiation \cite{1968ApJ...153....1P,RubinoMartin:2006ug,Chluba:2008aw,2009A&A...503..345C,2009RvMA...21....1S} can contribute energy into or extract energy from the photon-baryon plasma leading to spectral distortion in the early universe. In this work, we  study the spectral distortion caused by dissipation of acoustic wave (Silk Damping) for different dark matter models. A detailed description of that process is given in next section.

\subsection{Energy released due to Silk Damping} \label{heat}

At $z \ga 1100$, the photon and baryons are tightly coupled to each other via Compton scattering and behave like a single fluid. Adiabatic perturbations in this fluid behave like standing waves inside the sound horizon $r_s= c_s\eta$, where $c_s = c/\sqrt{3(1+R)}$ is the speed of sound in the plasma and $R = 3\rho_b/4\rho_{\gamma}$. At scales much smaller than the sound horizon, photon diffusion causes damping of density perturbation and bulk motion of this fluid \cite{1968ApJ...151..459S}. This process (Silk damping)  can be modelled by expanding the evolution of perturbations in the coupled photon-baryon fluid to second order in the mean free path of photons \cite{1970ApJ...162..815P,1983MNRAS.202.1169K}  or as the dissipation of energy of sounds waves owing to radiative viscosity and also on thermal conduction at late times \cite{1971ApJ...168..175W}. Silk damping causes the injection of entropy into the thermal plasma. 

As this is a continuous energy injection process, it can lead to $\mu$-type, $i$-type or $y$-type distortion depending on the era when the energy was injected. In the redshift range, $10^6 \ga  z \ga 10^3$, the structures  corresponding to  scales $k \simeq 0.3\hbox{--}10^4 \, \rm Mpc^{-1} $ are completely wiped out due to this damping. Thus, conversely,  one possible way to study and constrain the initial power spectrum at these scales is by observing the spectral distortion that is imprinted in the CMB spectrum due to this process. It should be noted that this is the only known probe of linear structures for such a wide range of small scales.  In comparison, the observed  CMB temperature anisotropies
from Planck  probe scales $k \la 0.1 \, {\rm Mpc^{-1}}$ \cite{Ade:2015xua} and the smallest scales probed by galaxy clustering data correspond to nearly linear scales at the present: $k \la 0.1 \, {\rm Mpc^{-1}}$ (e.g. \cite{Beutler:2016ixs}). 

The damping of adiabatic perturbations at small scales has been well studied in the literature. Recently a precise calculation of $\mu$- and $y$-type distortion due to Silk damping has been performed using the second order perturbation theory\cite{Chluba:2012gq}. It has been shown that tight-coupling approximation provides 
a good approximation for modelling the Silk damping in the pre-recombination era.
In this approximation, the source function for heating due to Silk damping is given by  \cite{Chluba:2012gq}\footnote{the source
function given in Eq.~(\ref{eq2.1}) can be understood as arising from radiative viscosity. In this case the energy pumped into the thermal plasma is proportion to the   square of the product of the photon mean free path with the velocity shear field (e.g. \cite{1977NCimR...7..277D}) which is proportional to $k^2 \Theta_1^2/\dot{\tau_c}^2$.  Alternatively, the dissipation of the energy can be modelled in terms of photon monopole \cite{Hu:1994bz}. Both these approaches give results within a factor of 3/4 of each other \cite{Chluba:2012gq}. While numerically computing CMB spectral distortion, we  use both the methods and find reasonable agreement}: 
\begin{equation}
\label{eq2.1}
S_{SD}(k,\eta) \simeq \frac{k^2}{\dot{\tau_c}^2}\left[\frac{R^2}{1+R}+\frac{16}{15}\right]|\Theta_{1}(k,\eta)|^2Y_{SZ}
\end{equation}
Where $\Theta_1(k,\eta)$ is the CMB dipole anisotropy, $\dot{\tau_c} = c n_e\sigma_T a$ is the derivative of Compton scattering optical depth with respect to the conformal time and $Y_{SZ}$ is the frequency dependent function representing the $y$-distortion defined in Eq.~\ref{eq1.8}. The first term in the square bracket comes from heat conduction and the second one is due to radiative  viscosity. At high redshift, the radiative  viscosity dominates as $R \rightarrow 0$.  As any time, the average  source function can be obtained  by integrating over all wavenumbers:
\begin{equation}
\label{eq2.1p} 
\langle S_{SD} \rangle(\eta) = \int {d^3k \over (2\pi)^3} S_{SD}(k,\eta)
\end{equation}
To obtain effective heating rate $\langle S_{SD} \rangle$  needs to be multiplied by $\dot{\tau_c}$ and  integrated over all frequencies. The integration over frequencies, $\int x^3 Y_{SZ} dx$,  yields $4\rho_{\gamma}$. This result provides the heating rate as a function of conformal time which is converted to a function of redshift by dividing it by $H(1+z)$. Thus, the final expression of heating rate is given by
\begin{equation}
\label{eq2.5}
\frac{1}{a^4\rho_{\gamma}}\frac{da^4Q}{dz} \approx \frac{4\dot{\tau_c} \langle S_{SD} \rangle}{H(1+z)}
\end{equation}
The $\mu$ and $y$-parameters are related to  heating rates as follows \cite{Chluba:2012gq,1993PhRvD..48..485H} 
 \begin{equation}
 \label{eq2.6}
  \mu = 1.4 \int_{z_{\mu}}^{\infty} \exp(-[z/z_{\mu}]^{5/2})\frac{dz}{a^4\rho_{\gamma}}\frac{da^4Q}{dz}
 \end{equation}
 and
 \begin{equation}
 \label{eq2.7}
 y = \frac{1}{4} \int_{z_{dec}}^{z_i}\frac{dz}{a^4\rho_{\gamma}}\frac{da^4Q}{dz}
 \end{equation}
 $z_{dec} \simeq  1100$ and $z_i \simeq 1.5 \times 10^4$  are the redshift of decoupling and the redshift denoting the end of $i$-distortion era, respectively.

\section{Spectral distortion and dark matter models} \label{cosmology}

In this section, we discuss how altering the dark matter model impacts the generation of entropy owing to Silk damping in the pre-recombination era. 
As noted in the previous section, the dynamics of this energy pumping can be captured by the time evolution of the photon dipole, $\Theta_1(k,\eta)$.

We discuss  here how the photon dipole is altered when the dark matter model is changed. We use Newtonian conformal gauge for motivating our discussion as the underlying physics is more transparent in this gauge. In the tight-coupling approximation, the photon dipole can be expressed as (e.g. \cite{2003moco.book.....D}):
\begin{equation}
\Theta_1(k,\eta) = \left[{1\over \sqrt{3}}\left(\Theta_0(k,0)+\Phi(k,0)\right)\sin(kc_s\eta)-{k\over3}\int_0^\eta d\eta'\left(\Phi(k,\eta')-\Psi(k,\eta')\right)\cos(kc_s\eta-kc_s\eta')\right] \exp(-k^2/k_d^2)
\label{eq:theta1}
\end{equation}
Here $\Theta_0(k,\eta)$ is the photon monopole and $\Phi(k,\eta)$ and $\Psi(k,\eta)$ are the two 
gravitational potentials in the Newtonian gauge.
If the small contribution from neutrinos  is neglected, $\Psi = -\Phi$. \footnote{$\Psi=-(1+2/5R_\nu)\Phi$ with $R_\nu = \rho_\nu/(\rho_\nu+\rho_\gamma)$;  $R_\nu = 0.41$ for three massless neutrino degrees of freedom. Two of the models we consider here can alter  $R_\nu$. This  means that the first term 
in Eq.~(\ref{eq:theta1}), which corresponds to initial conditions, can also
be used to distinguish between different dark matter models \cite{Chluba:2012gq}.   In WDM models, the dark matter particle is initially
relativistic and therefore would increase $R_\nu$. However, even during this
phase the contribution of this particle to neutrino relativistic degrees of 
freedom is negligible  as compared to the standard model neutrino;  for $m_{\rm wdm} = 1 \, \rm keV$, the
particle contributes less than 1\% of the energy density of a standard model
neutrino. In LFDM models, a tiny fraction of neutrino energy density is used
to create the dark matter particle during a phase transition (e.g. \cite{Sarkar:2014bca}). For both these models, the effect of initial conditions is generally
negligible. We take into account these effects in our detailed modelling using 
\texttt{CMBFAST} but drop it in this sub-section to uncover the main determinants of 
the impact of dark matter models on the photon dipole.}
 $\Theta_0(k,0) = 0.5 \Phi(k,0)$ and $c_s = 1/\sqrt{3(1+R)}$.  In Eq~(\ref{eq:theta1}) we have neglected $R =3\rho_B/4\rho_\gamma$ and its evolution in
the pre-factors of the equation. \footnote{For the charged decay model, $R = 3(\rho_B+\rho_{\rm dm})/(4\rho_\gamma)$ at early times as the dark matter particle
is charged and tightly coupled to the baryon-photon fluid. For this 
model, this is an additional effect that determines the evolution of $\Theta_1(k,\eta)$. In the results presented in this section, we take into account 
this effect using \texttt{CMBFAST}.} $k_d(\eta)$ corresponds to the scale that
undergoes Silk damping at any time:
\begin{equation}
k_d^{-2}(\eta) \simeq \int_0^\eta d\eta {1\over \dot\tau}\left({R^2 \over 6(1+R)^2} +{4 \over 27(1+R)} \right )
\label{eq:silkdamsca}
\end{equation} 
Here $\dot\tau = n_e\sigma_T a$. Under these approximations, the evolution of $\Theta_1(k,\eta)$ can be 
completely determined by the Newtonian potential  $\Phi(k,\eta)$ and the Silk damping scale  $k_d(\eta)$. As discussed in the next section, the dissipation of energy at a given scale $k$ occurs predominantly at a time $\eta_d$ such that $k = k_d(\eta_d)$. As we argue below, the main effect we seek depends on the time evolution of 
$\Phi(k,\eta)$ in the time range, $1/k < \eta < 1/k_d(\eta_d)$, i.e. from the horizon
entry of a scale to the time at which the  mechanical energy at this scale    dissipates. 

All the models we consider in this paper are based on altering the nature of CDM with an aim to suppress power at small scales. While these models impact the radiation content of the universe in the early universe (e.g. WDM
model is based on a particle of mass $m_{\rm wdm} \simeq 1 \, \rm kev$ which is relativistic in the early universe) they all leave unchanged the matter-radiation equality epoch. This implies that their impact is proportional to the ratio of dark matter to the radiation energy density,  $\rho_{\rm dm}/\rho_r$, which is much smaller than unity at early times. This means that the impact of changing the dark matter change is essentially captured by the second term of Eq.~(\ref{eq:theta1}), 
which depends on the evolution of the Newtonian potential $\Phi(k,\eta)$. 

\begin{figure}
\begin{subfigure}{.5\textwidth}
  \centering
  \includegraphics[width=1.0 \linewidth ]{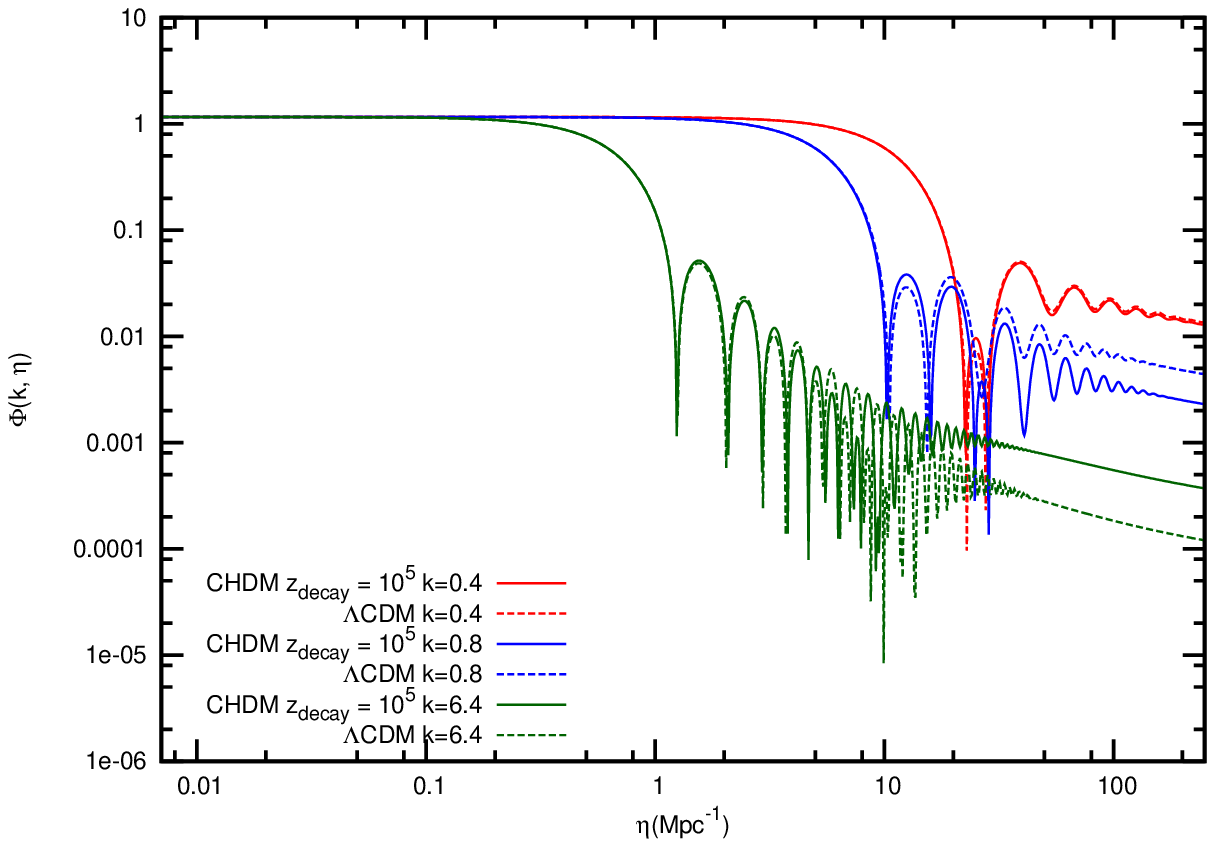}
  \label{ea}
\end{subfigure}%
\begin{subfigure}{.5\textwidth}
  \centering
  \includegraphics[width=1.0 \linewidth ]{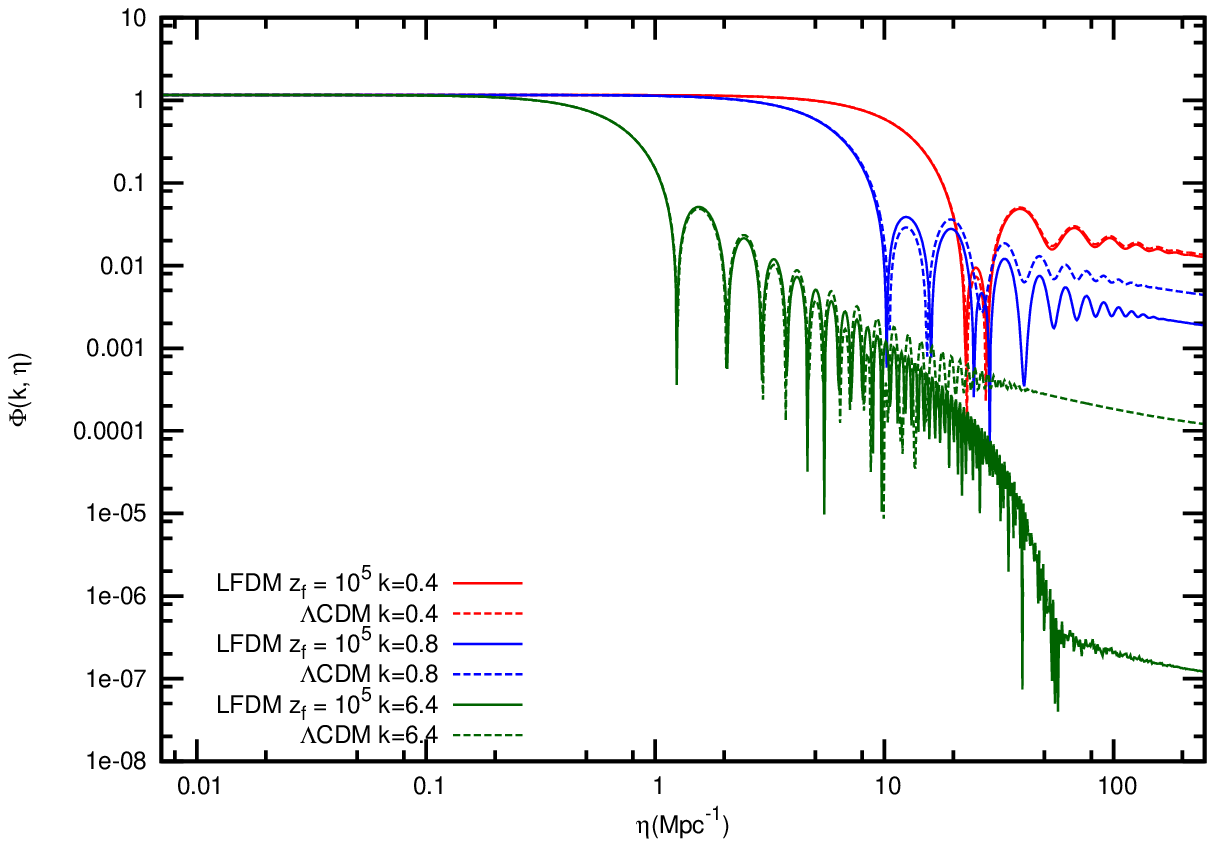}
  \label{feb}
\end{subfigure} 
\caption{The evolution of $\Phi(k,\eta)$, for $k=0.4,0.8,6.4 \rm {Mpc^{-1}}$, for CHDM (Left panel) and LFDM(Right panel) models  are compared to the  $\Lambda$CDM model. In each subfigure the solid lines stand for the non-standard model and the dotted lines for  $\Lambda$CDM model.}
\label{phi}
\end{figure}

In Figure~\ref{phi} we show the evolution of $\Phi(k,\eta)$ for two of the models 
we consider here for a range of wavenumbers $k$. The Newtonian potential 
$\Phi(k,\eta)$ is computed from \texttt{CMBFAST} code, which is written in Synchronous gauge,  using the following transformation:
\begin{equation}
\Phi(k,\eta) = \beta(k,\eta) - {\dot a\over a  k^2} \left (\dot h(k,\eta)+6\dot\beta(k,\eta) \right )
\end{equation}
Here $h$ and $\beta$ are the potentials in Synchronous Gauge (Eq.~18 of \cite{Ma:1995ey}). Some notable features of the comparison between the models we consider and the $\Lambda$CDM models are: (a) the difference between the potentials
is negligible of for large scales (small $k$). This is expected as the models
we consider agree with the $\Lambda$CDM model on  large scales, (b)  As the scale gets smaller,  the  difference 
between alternative DM models and the $\Lambda$CDM model becomes  more significant at  larger $\eta$. We discuss the 
nature of this deviation below.

{\bf LFDM model}: In this model, the CDM forms at a redshift $z = z_f$ from a tiny fraction of relativistic neutrinos due to a phase transition. It inherits the density and velocity perturbations of the neutrino component, resulting in a sharp reduction in matter power for scales $k > k_{\rm lfdm}$, where $k_{\rm lfdm}$ is the scale that enters the horizon at $z = z_f$.

To understand Figure~\ref{phi} we  consider scales smaller and 
larger than $k_{\rm lfdm}$: (a) $k < k_{\rm lfdm}$:  these scales are outside the horizon when the dark matter forms. These scales evolve outside the horizon in a purely radiation dominated universe. The perturbations at these scales are not affected by LFDM physics except in determining the initial conditions which are not necessary for the following reasons: the photon perturbations outside the horizon are constant and are set by the potential. For scales outside the horizon, the potential changes by a factor 9/10 in making a transition from radiation to the matter-dominated era (e.g. \cite{2003moco.book.....D} and references therein). If the era of LFDM lies deep inside RD era, 
the change is negligible. This explains the large scale behaviour of potential in Figure~\ref{phi}.  

 (b) $k > k_{\rm lfdm}$: these scales enter the horizon 
before the formation of the cold dark matter.  As $z_{\rm lfdm} \gg z_{\rm eq}$ these scales  evolve sub-horizon  in the radiation dominated era. In radiation dominated era, the potential $\Phi(k,\eta)$ decays as $(k\eta)^{-2}$ for $k\eta \gg 1$ if perturbations in radiation  determine the evolution of the potential (e.g. \cite{2003moco.book.....D}). At sub-horizon scales, neutrino perturbation decay exponentially while photon density perturbations equal baryon perturbations which oscillate with nearly constant amplitude for $\eta < \eta_d$. The CDM perturbations, on the other hand, grow logarithmically during this phase.  This means matter perturbations can determine the evolution of the potential $\Phi(k,\eta)$ even in deep radiation dominated era. This behaviour is seen the evolution of the potential for $k = 0.8$ in Figure~\ref{phi} as flattening of the potential for larger $\eta$. For LFDM models, the potential decay is sharper because of the CDM forms late. This behaviour along with other features for $k\eta > 1$ seen in Figure~\ref{phi} explains how the evolution of $\Phi$ in  LFDM models differs from the $\Lambda$CDM model. It is important to note that this difference is most pronounced for scales
that enter the horizon at $z \simeq z_f$. For even smaller scales, the effect of matter perturbations is less important as these scales enter the horizon when $\rho_{\rm dm}/\rho_r$ is lower and therefore matter perturbations have a smaller impact on their evolution. Equivalently, it is seen from Eq.~(\ref{eq:theta1}) that the photon dipole depends on the integral of the potential and as the potential decays inside the horizon, most of the contribution comes from epochs close to the horizon entry. It is also seen in the evolution of the potential for $k = 6.4 \, \rm Mpc^{-1}$. The potential deviates insignificantly from the $\Lambda$CDM model for $k\eta \gtrsim 1$ but shows sharp deviation at later times. The evolution of perturbations at these scales (from $\eta \ll 1$ to $\eta= \eta_d$) correspond roughly to the case of no dark matter at all time.  The potential for this case falls exponentially for large $\eta$ which occurs because the only source of potential in the LFDM model is photon and neutrino perturbations which both decay exponentially for $\eta > \eta_d$. It is of interest to note that 
the potential at $k = k_d$ is dominated by matter perturbations for $z \lesssim 10^5$.

For the LFDM model shown in Figure~\ref{phi}, the scales of interest for measuring the deviation from the $\Lambda$CDM model lie in the range $0.6\hbox{--}4 \, \rm Mpc^{-1}$. These scales enter the horizon before, but not significantly before,  the era of matter formation and the perturbations at these scales suffer Silk damping  after $z \simeq 10^4$.

For models such as WDM and ULA, the preceding discussion is directly applicable. We briefly discuss the charged decaying particle case below.

{\bf Charged Decaying particle}:
For charged decay model, the variation in the dynamics of $\Phi(k,\eta)$ closely follows the previous discussion for LFDM model: $\Phi(k,\eta)$ nearly follows the potential for $\Lambda$CDM model at large scales. At very small scales the effect is
suppressed because the matter density is small when they enter the horizon (Figure~\ref{phi}). The main impact is captured by intermediate scales that enter the horizon around
but before the time of decay. 

Some of the salient differences between the two cases can be seen in Figure~\ref{phi}. For the Charged decaying particle model, the potential can exceed the 
$\Lambda$CDM values for a range of wavenumbers, e.g. $k=6.4 \, \rm Mpc^{-1}$ in the figure. 

It follows from our discussion that the impact of alternative DM  models on $\Phi(k,\eta)$  declines if the transition (decay) redshift
moves deeper into the radiation dominated era. We note that the tight-coupling approximation while capturing the essential physics, tends to break down close to the recombination era; this is relevant for the computation of $y$-distortion \cite{2012A&A...543A.136K,Chluba:2012gq}. However,  this approximation allows us to compare our results in two different gauges and identify the main determinants of altering CMB distortion for alternative dark matter models.

\begin{figure}
\begin{subfigure}{.3\textwidth}
  \centering
  \includegraphics[width=1.0 \linewidth ]{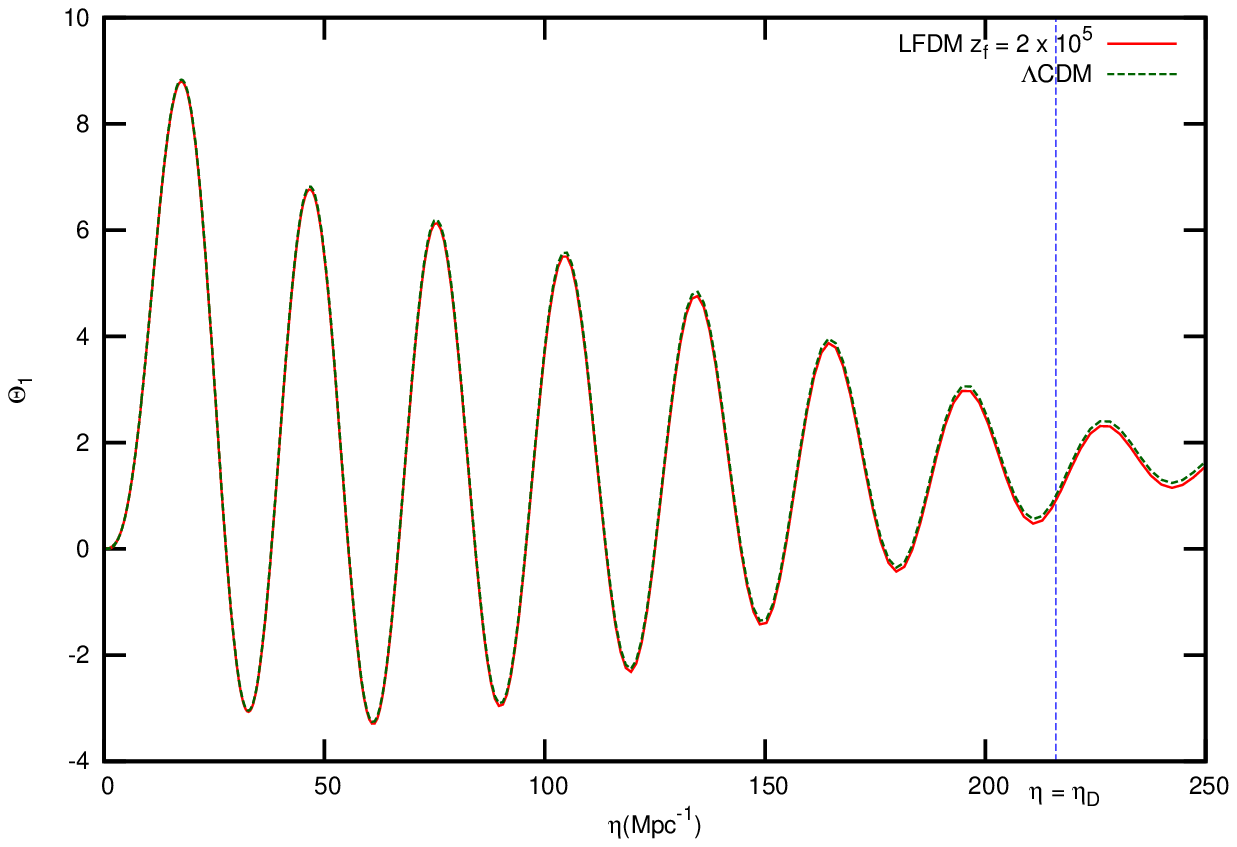}
  \label{fig:ta}
\end{subfigure}%
\begin{subfigure}{.3\textwidth}
  \centering
  \includegraphics[width=1.0 \linewidth ]{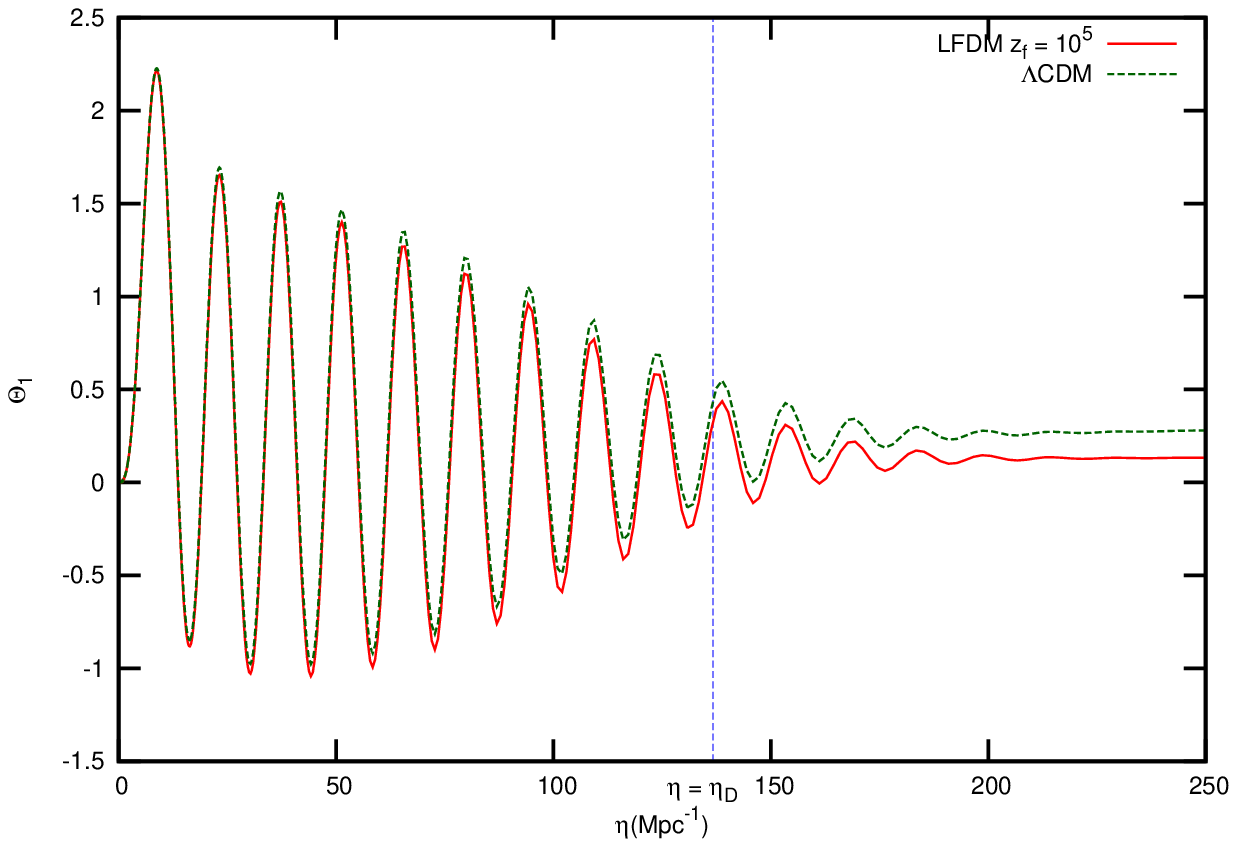}
  \label{fig:tb}
\end{subfigure} 
\begin{subfigure}{.3\textwidth}
  \centering
  \includegraphics[width=1.0 \linewidth ]{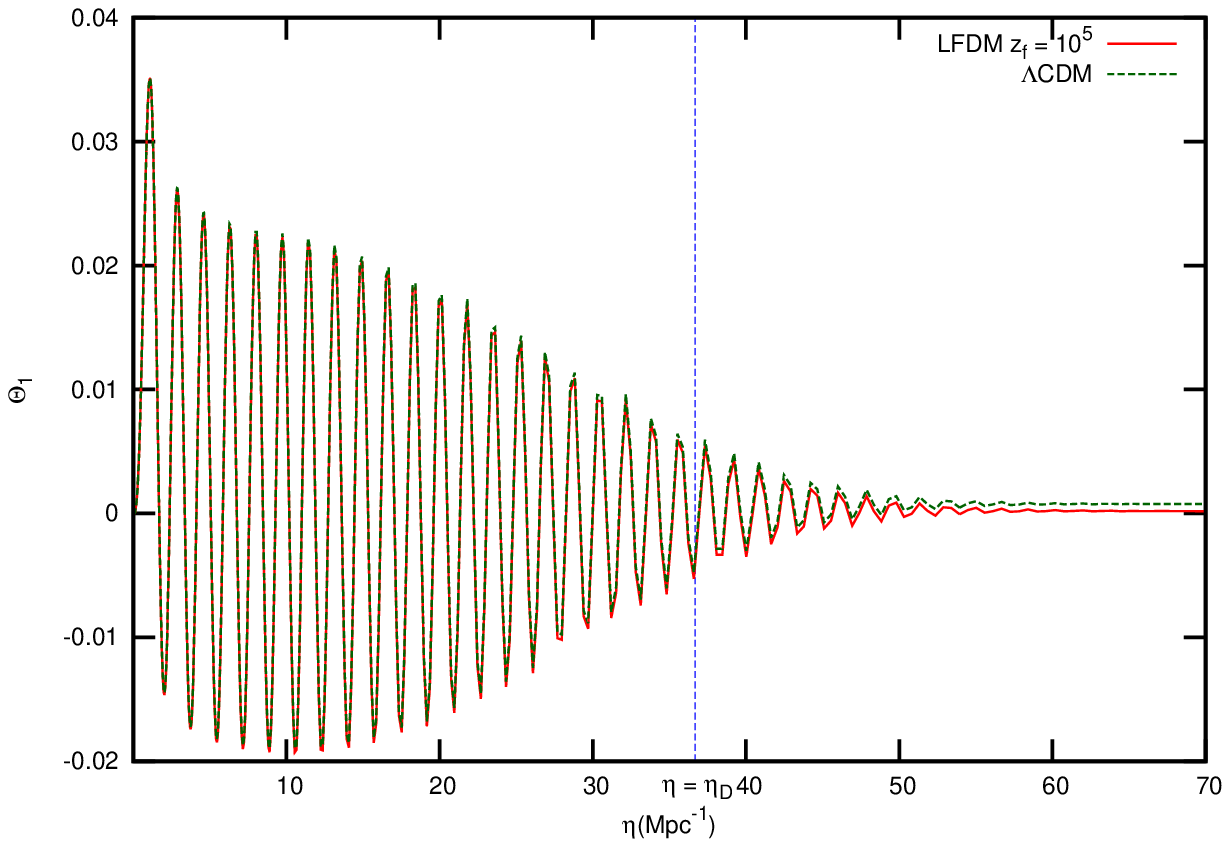}
  \label{fig:tc}
\end{subfigure} \\
\begin{subfigure}{.3\textwidth}
  \centering
  \includegraphics[width=1.0 \linewidth ]{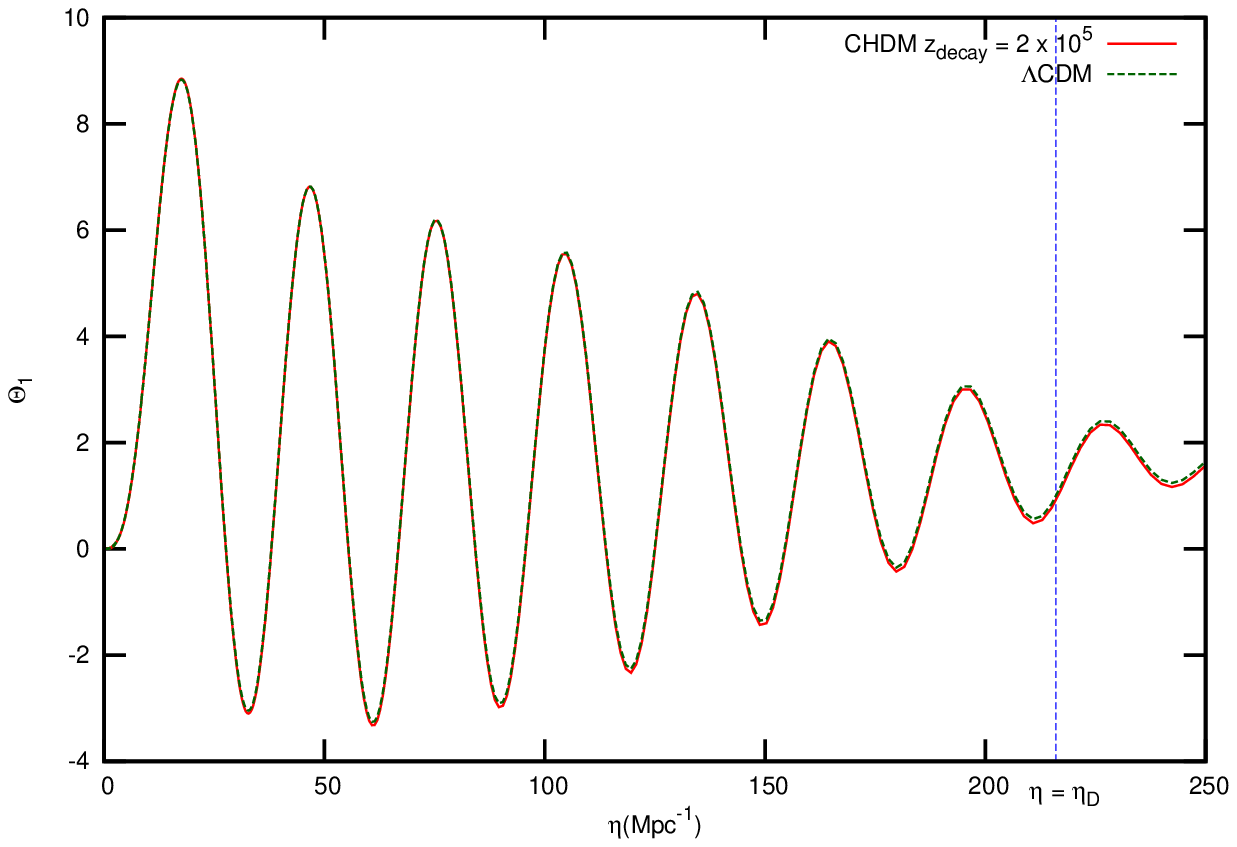}
  \label{fig:td}
\end{subfigure}%
\begin{subfigure}{.3\textwidth}
  \centering
  \includegraphics[width=1.0 \linewidth ]{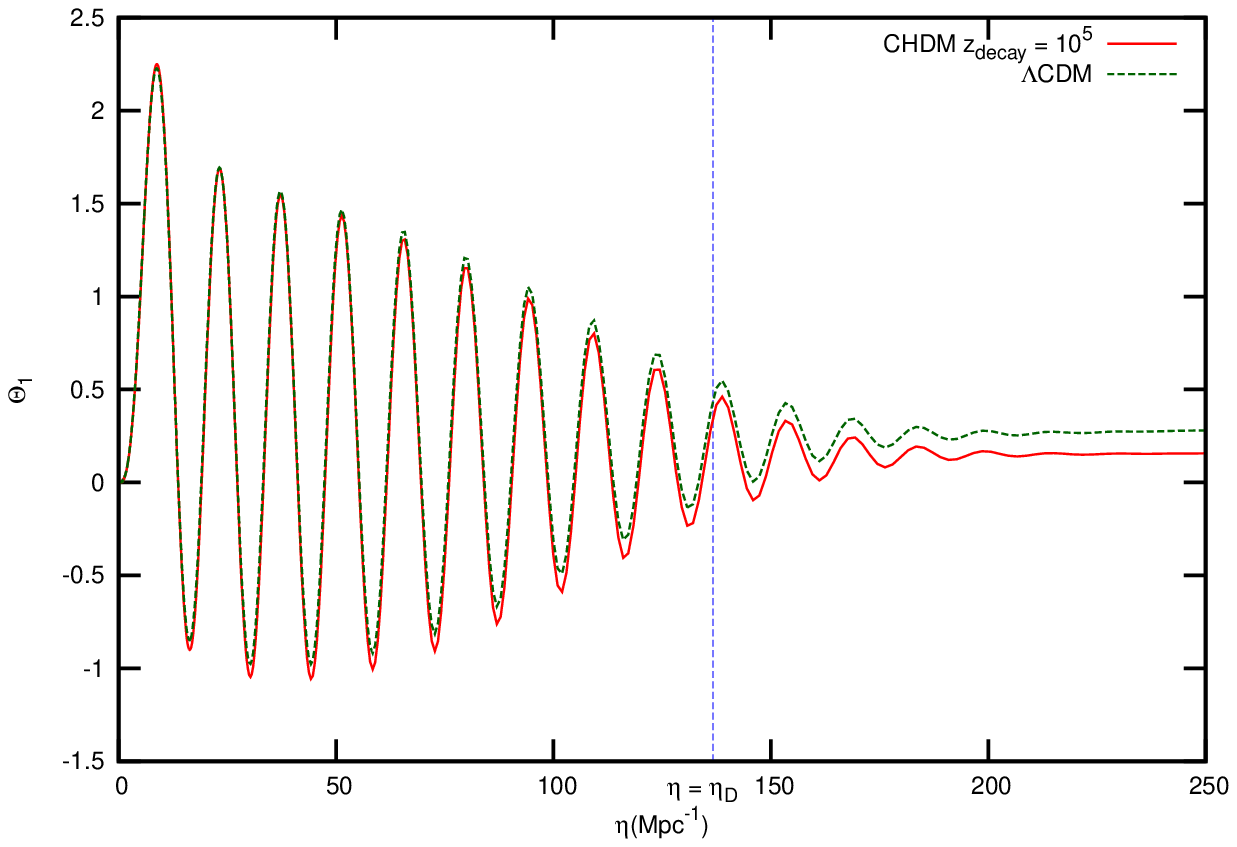}
  \label{fig:te}
\end{subfigure} 
\begin{subfigure}{.3\textwidth}
  \centering
  \includegraphics[width=1.0 \linewidth ]{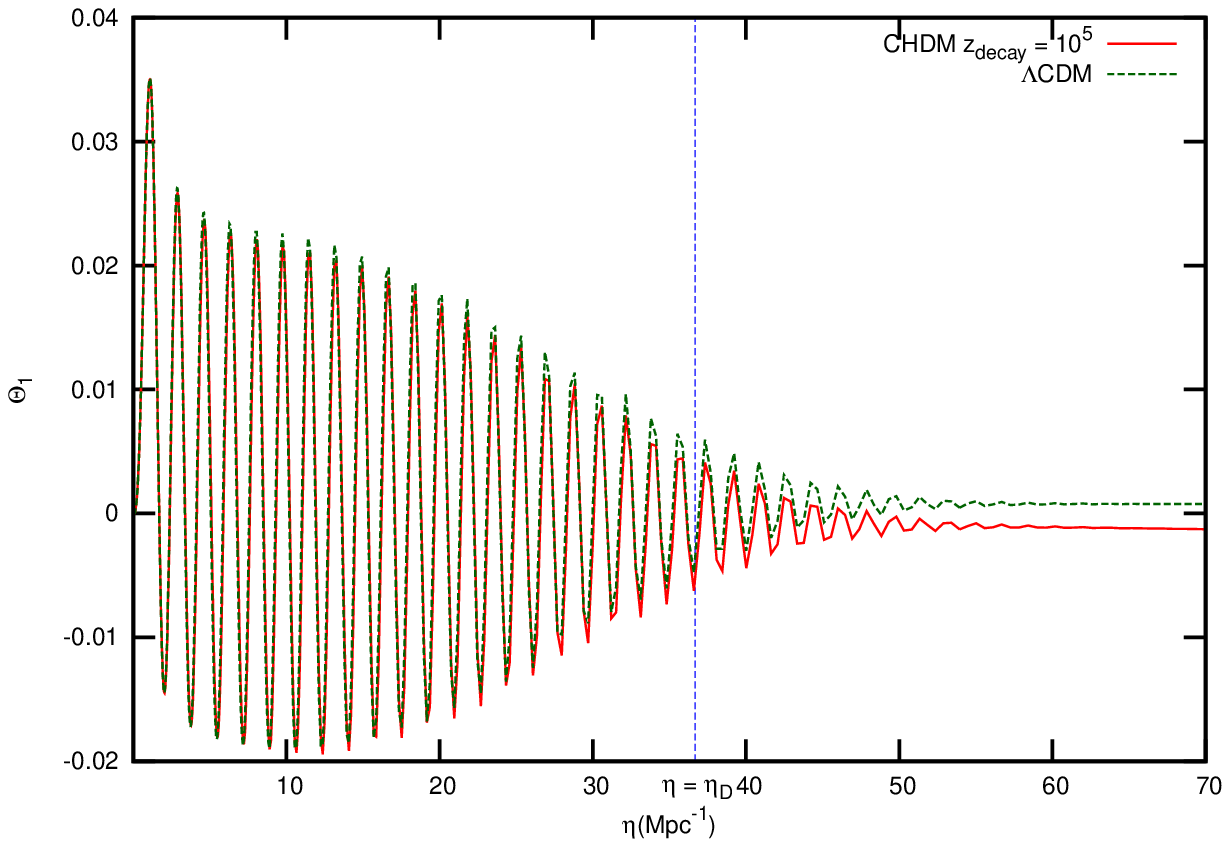}
  \label{fig:tf}
\end{subfigure} \\
\caption{The evolution of $\Theta_1(k,\eta)$ for $k =0.4,0.8,6.4 \rm{Mpc^{-1}}$ for CHDM (lower panel) and LFDM (upper panel) for $z_f = z_{\rm decay} =  10^5$. In each plot we also show the evolution of the corresponding scale for $\Lambda$CDM model. In each plot the vertical dotted line marks the time when $k = k_d$ (Eq.~(\ref{eq:silkdamsca})) or  that particular scale enters the Silk damping regime.}
\label{theta}
\end{figure}

\begin{figure}
\centering
\includegraphics[height = 2.5 in, width = 4.5 in]{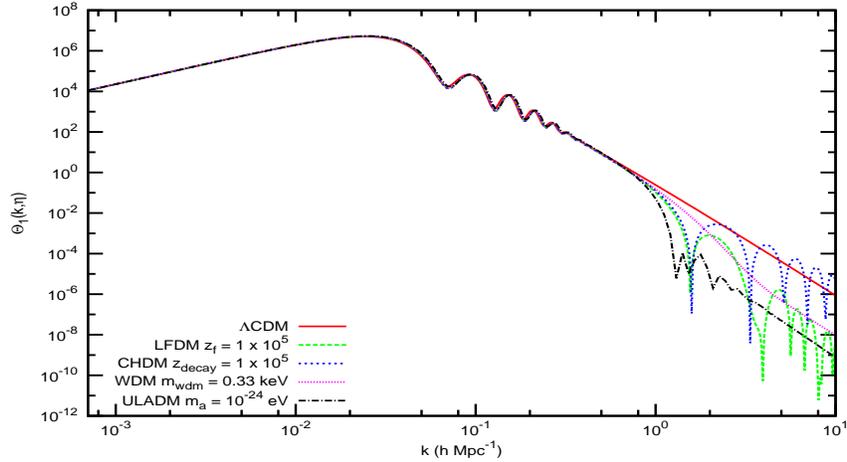}
\caption{CMB dipole transfer function of the four dark matter candidates considered in this work along with $\Lambda$CDM. The specifications of the models are same as that of Figure-\ref{powerspec}. }
\label{power1}
\end{figure}

In Figure~\ref{theta} we show the dynamics of $\Theta_1(k,\eta)$ for a range of alternative DM models. We compute $\Theta_1(k,\eta)$ directly from \texttt{CMBFAST} code but also verify that Eq.~(\ref{eq:theta1}), which shows that $\Theta_1(k,\eta)$ depends on the history of the variation of $\Phi(k,\eta)$, explains the main difference expected 
for alternative DM models. We establish it by running a smaller set of equations for density and velocity  perturbations in photon-baryon fluid and  dark matter along with the evolution of $\Phi(k,\eta)$ in conformal Newtonian gauge \cite{Ma:1995ey}.

In Figure~\ref{power1}, we show the values of $\Theta_1(k,\eta)$ at $z = 1000$ of different models at the same redshift. All  the models shown correspond roughly to cases where the decrement 
in the power spectrum occurs at $k \simeq  0.3 h \, \rm   Mpc^{-1}$. 

\textcolor{black}{The preceding discussion allows us to establish that the main impact of alternative dark matter models is to alter $y$-distortion and not $\mu$- or $i$-distortion. The $\mu$-distortion era occurs in the redshift range $2\times 10^5 \lesssim  z \lesssim 2 \times 10^6$. In this era the scales that dissipate their energy and therefore cause the CMB distortion lie in the range $500  \, {\rm Mpc^{-1}} \lesssim k  \lesssim  14000 \, {\rm Mpc^{-1}}$. In the range of redshifts of interest, the matter power for alternative models at these scales is highly suppressed as compared to the $\Lambda$CDM model, as seen in Figure~\ref{transfer}. However, as argued above, this does not have a significant impact on $\Theta_1$ because it depends on the history of the evolution of the potential $\Phi(k, \eta)$ (Eq.~(\ref{eq:theta1})). As the potential decays after horizon entry (Figure~\ref{phi}), the photon dipole is governed by the initial condition of the potential and its evolution closer to the time of horizon entry of the scale ($\eta = \eta_e$) rather than its value at $\eta = \eta_d$. The scale $k \simeq  500 \, \rm Mpc^{-1}$ (which decays at the end of the $\mu$-distortion  era)  enters the horizon at $z \simeq 2 \times 10^8$. As the ratio of radiation to matter energy density at this redshift $\rho_{\rm dm}/\rho_r \simeq 10^{-5}$, the potential is insensitive to even a substantial change in the matter power at this scale at early times. The initial conditions for different models are not the same as the $\Lambda$CDM model.   However, as discussed above, requiring these models to agree with large scale observations forces the impact of the change in initial conditions on the potential to be negligible.}

\textcolor{black}{This also implies that the dynamics of scales that determine $\mu$-distortion of CMB are minimally affected by a change in the dark matter model. As we showed above, the main effect in the photon dipole is caused by scales in the range $0.6\hbox{--}4 \, \rm Mpc^{-1}$, this also means the impact of alternative dark matter on the $i$-distortion era is also negligible. We test this in our study by computing the photon dipole for scales in the range   $k < 500 \, \rm Mpc^{-1}$  for alternative models numerically and using an analytic approximation for even smaller scales to discern the impact of initial conditions (the first term of Eq.~(\ref{eq:theta1})). For all the models we consider, the $\mu$- and $i$-distortion is less than $0.1 \%$.}

\section{Results} \label{result}

Using Eqs.~(\ref{eq:theta1}) and~(\ref{eq2.1}) one can analytically show that the dissipation at any given time is dominated by scales such that $k \simeq k_d$. Such an estimate is based on approximating $\Theta_1(k,\eta)$ by  the first term of Eq.~(\ref{eq:theta1}) (e.g. \cite{Chluba:2012gq}). This gives:
\begin{equation}
\label{eqsd5.1}
S_{SD}(k,\eta) \propto \frac{k^2}{\dot{\tau_c}^2}P_\phi(k,0) \sin^2(kc_s\eta) \exp(-k^2/k_d^2)Y_{SZ}
\end{equation}
Here $P_\Phi(k,0)$ is  the power spectrum of the  potential at the initial time. Using $P_\Phi(k,0) \propto k^{-4+n_s}$ (e.g. \cite{Chluba:2012gq,2003moco.book.....D}) where $n_s \simeq 1$ is the scalar spectral index and integrating over $k$ gives:
\begin{equation}
\label{eqsd5.2}
\langle S_{SD} \rangle \propto \int dk k^{n_s}  \sin^2(c_sk\eta) \exp(-k^2/k_d^2)
\end{equation}
This $k$ integral in the equation is dominated by $k\simeq k_d$ which means  the energy  dissipation at any given time is governed by this condition.  This result holds even when both the terms in Eq.~(\ref{eq:theta1}) are retained.\footnote{While the first term of Eq.~(\ref{eq:theta1}) gives a reasonable analytic estimate, the second term generally dominates the first term even for the  $\Lambda$CDM model. The factor  proportional to $\sin(c_sk\eta)$ in the second term dominates over the first term and cancels it but the remainder has the form of the first term and has a similar order of magnitude. The factor proportional to $\cos(c_sk\eta)$ in the second term is generally sub-dominant. These two factors contain information of the evolution of potential which is germane to our work.} 

\begin{figure}
\begin{subfigure}{.5\textwidth}
  \centering
  \includegraphics[width=1.0 \linewidth ]{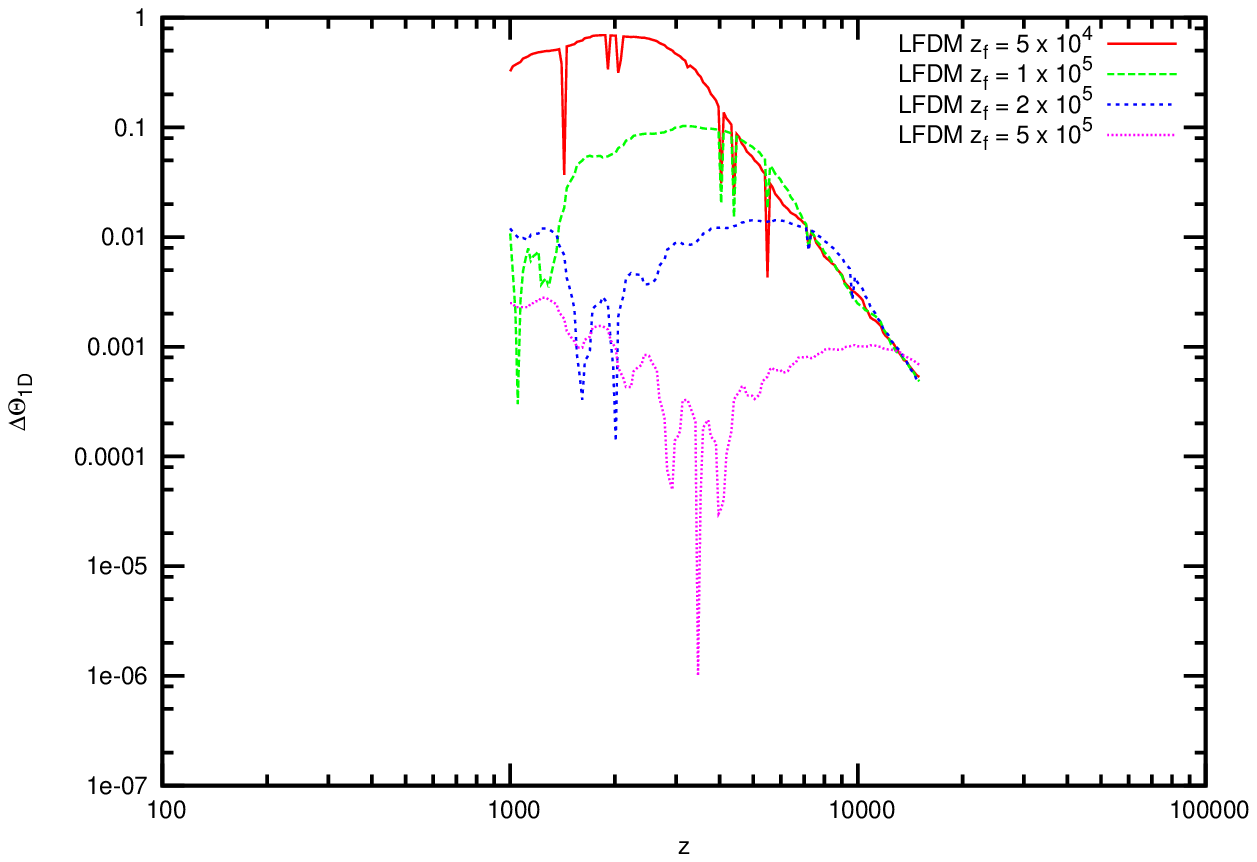}
  \label{fig:3a}
\end{subfigure}%
\begin{subfigure}{.5\textwidth}
  \centering
  \includegraphics[width=1.0 \linewidth ]{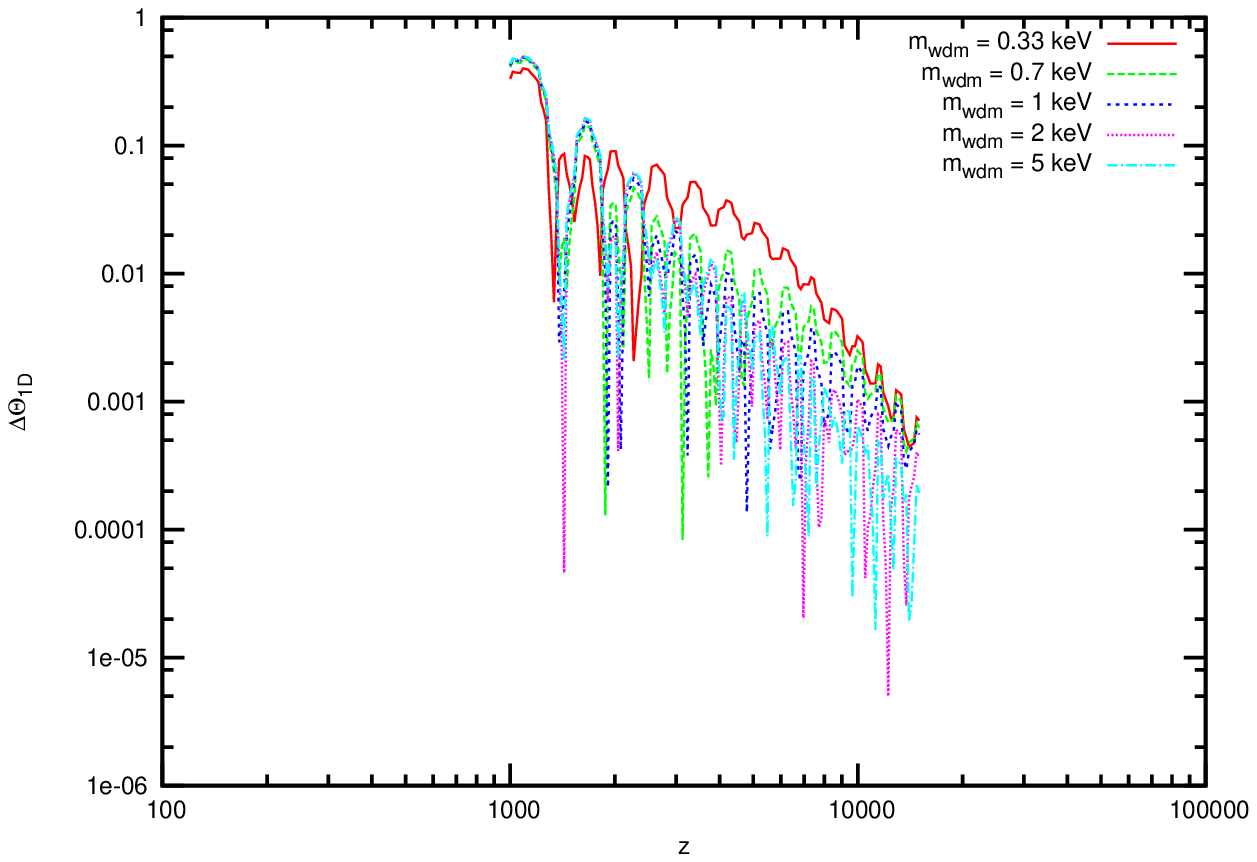}
  \label{fig:3b}
\end{subfigure} \\
\begin{subfigure}{.5\textwidth}
  \centering
  \includegraphics[width=1.0 \linewidth ]{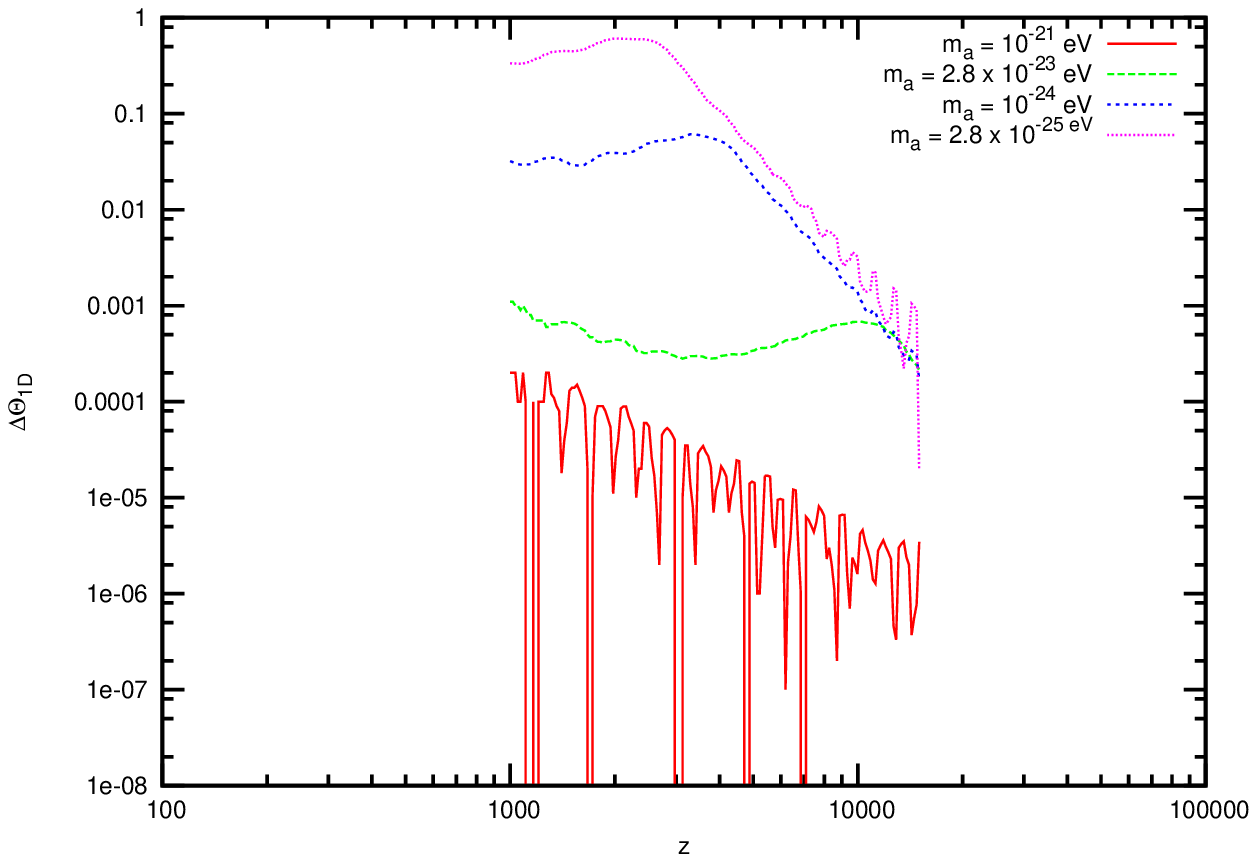}
  \label{fig:3c}
\end{subfigure}%
\begin{subfigure}{.5\textwidth}
  \centering
  \includegraphics[width=1.0 \linewidth ]{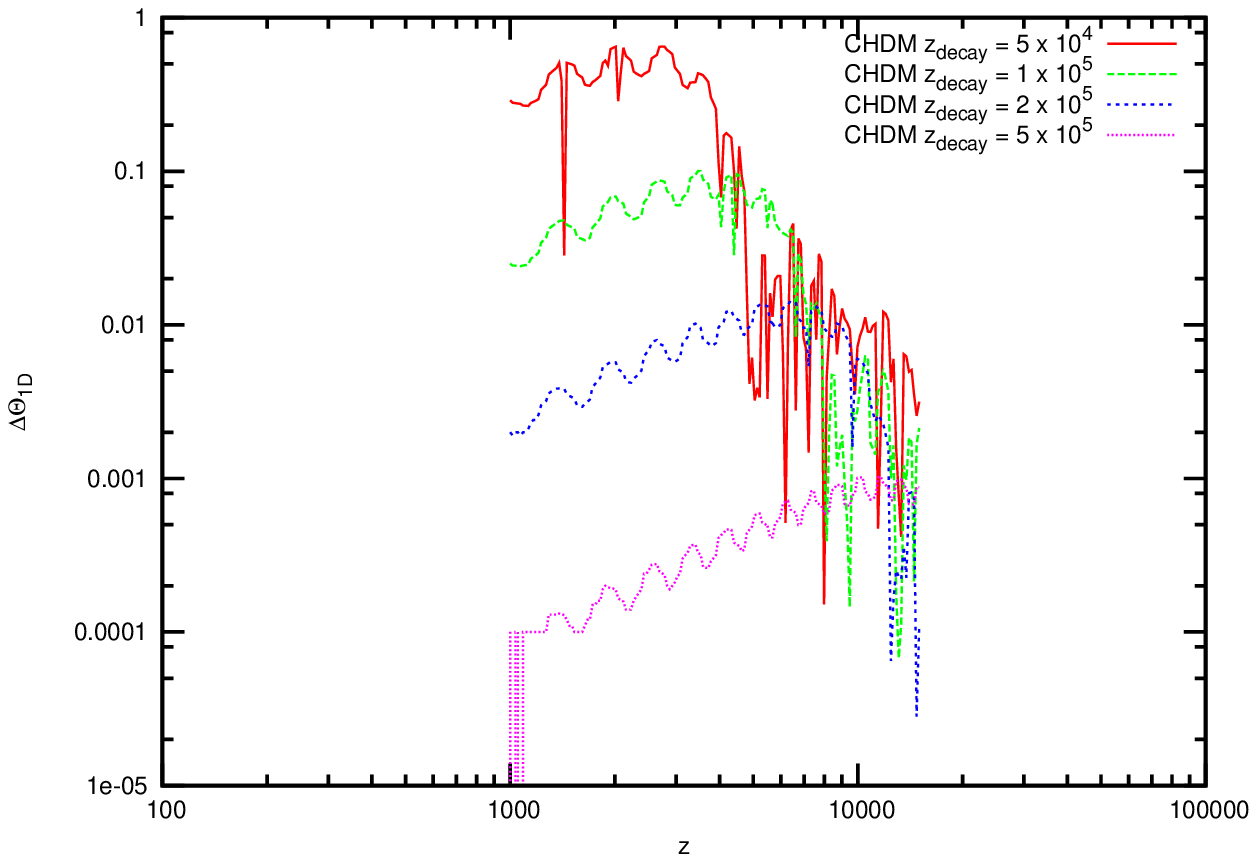}
  \label{fig:3d}
\end{subfigure} \\
\caption{The difference between $\Theta_{1\rm{D}}$ for dark matter models considered in this work and  $\Lambda$CDM model: $\Delta\Theta_{1\rm{D}}$. Clockwise from top-left $\Delta\Theta_{1\rm{D}}$ for LFDM, WDM, charged particle decay dark matter and ULADM are plotted, respectively.}
\label{fig3}
\end{figure}

This means that the dissipation is dominated by photon dipole at $k\simeq k_d$ at any time $\eta$,  $\Theta_{1\rm{D}} \equiv \Theta_1(k_d,\eta)$. In Figure~\ref{fig3} we show the redshift evolution of $\Delta\Theta_{1\rm{D}}$,  the difference between $\Theta_{1\rm{D}}$ for different dark matter models and that of $\Lambda$CDM. The set of model parameters we have chosen is provided in Table~\ref{tab1}.  The difference decreases as the redshift of formation $z_f$ for LFDM, and the decay redshift $z_{\rm decay}$ for CHDM and masses of WDM and Axion are increased. One further point to note is that the difference  $\Delta\Theta_{1\rm{D}}$ for alternative models increases as the redshift decreases. These results are in agreement with our discussion in the previous section which suggested that the alternative dark matter models yield substantial difference only during the $y$-distortion era.

\begin{table}
\begin{center}
 \begin{tabular}{|c|c|c|c|} 
 \hline
 Model     & Parameter     & $y \times 10^{9}$         &$\%$ difference of $y$ from $\Lambda$CDM     \\ [0.5ex] 
 \hline
 $\Lambda$CDM     & \cite{Ade:2015xua}     & $4.4180 $     &$0.0$    \\ [0.5ex]
 \hline
 \,     & $z_f=5 \times 10^4$ &  $3.8561 $    &$14.57$         \\ [0.5ex] \cline{2-4}
 \,     & $z_f=1 \times 10^5$ &  $4.1001 $    &$7.75$   \\ [0.5ex] \cline{2-4}
 LFDM  & $z_f=2 \times 10^5$ & $4.3037$      &$2.65$      \\ [0.5ex] \cline{2-4}
 \,		& $z_f=5 \times 10^5$ & $4.3959 $     &$0.50$     \\ [0.5ex] \cline{2-4}
 
 \hline
 \,     & $m_{\rm wdm}=0.33$ keV   & $4.2178$    &$4.74$   \\ [0.5ex] \cline{2-4}
 \,     & $m_{\rm wdm}=0.70$ keV    & $4.3105 $   &$2.49$    \\ [0.5ex] \cline{2-4}
 WDM    & $m_{\rm wdm}=1.00$ keV    & $4.3398 $   &$1.80$   \\ [0.5ex] \cline{2-4}
 \,     & $m_{\rm wdm}=2.00$ keV   & $4.3680 $    &$1.14$  \\ [0.5ex] \cline{2-4}
 \,     & $m_{\rm wdm}=5.00$ keV    & $4.3798 $   &$0.87$   \\ [0.5ex]
 \hline
 \,     & $z_{\rm decay}=5 \times 10^4$ & $3.8913 $    &$13.53$        \\ [0.5ex] \cline{2-4} 
 CHDM     & $z_{\rm decay}=1 \times 10^5$ & $4.1884 $  &$5.48$         \\ [0.5ex] \cline{2-4}
 \,  & $z_{\rm decay}=2 \times 10^5$ & $4.2945 $   &$2.87$  \\ [0.5ex] \cline{2-4}
 \,		& $z_{\rm decay}=5 \times 10^5$ & $4.4002 $  &$0.4$       \\ [0.5ex] \cline{2-4}
 \hline
 \,     & $m_a=2.8 \times 10^{-25}$ eV   & $3.8840 $  &$13.74$       \\ [0.5ex] \cline{2-4}
 \,     & $m_a=1.0 \times 10^{-24}$ eV   & $4.2812 $  &$3.19$       \\ [0.5ex] \cline{2-4}
 ULA DM    & $m_a=2.8 \times 10^{-23}$ eV     & $4.3990 $   &$0.43$      \\ [0.5ex] \cline{2-4}
 \, & $m_a=1.0 \times 10^{-21}$ eV     & $4.4177 $   &$6.8 \times 10^{-3}$     \\ [0.5ex] \cline{2-4}
 \hline
\end{tabular}
\end{center}
\caption{This table lists  the values of $y$-parameter for  alternative dark matter models and compared with $\Lambda$CDM model.} 
\label{tab1}
\end{table}

Using  Eqs.~(\ref{eq2.1}), (\ref{eq2.1p}) and~(\ref{eq2.7}),  we calculate $y$-parameter for different models; $\Theta_1(k,\eta)$ is normalized using the condition on mass dispersion at scale $R =8 \, \rm h^{-1} Mpc$ at the present epoch: $\sigma_8(\eta_0) = 0.8225$ \cite{Ade:2015xua}. The list of $y$-parameters for different models is provided in Table~\ref{tab1}. We note that the $y$-parameter
can vary by up to $10\%$ for a range of models.  This behaviour is in line  with the variation of  $\Delta\Theta_{1\rm{D}}$. The $y$-parameter approaches its value for $\Lambda$CDM model as $z_f$ for LFDM and $z_{\rm decay}$ for CHDM models are increased, and also  when the mass of WDM and axion is increased as expected from our discussion above. Furthermore, we can use Eq.~(\ref{eq1.8}) to determine the shape of distorted CMB spectrum.

\begin{figure}[H]
\begin{subfigure}{.5\textwidth}
  \centering
  \includegraphics[width=0.95 \linewidth ]{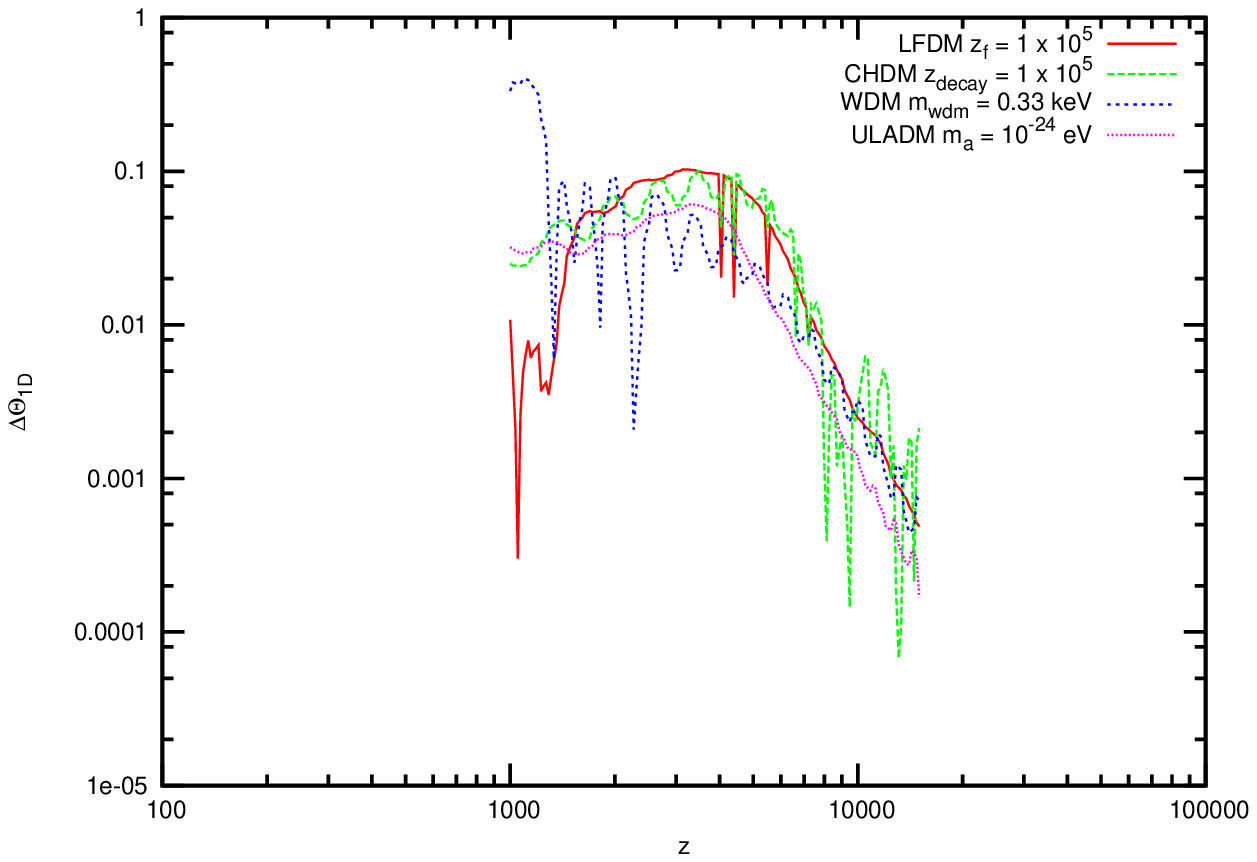}
  \label{fig:1a}
\end{subfigure}%
\begin{subfigure}{.5\textwidth}
  \centering
  \includegraphics[width=0.95 \linewidth ]{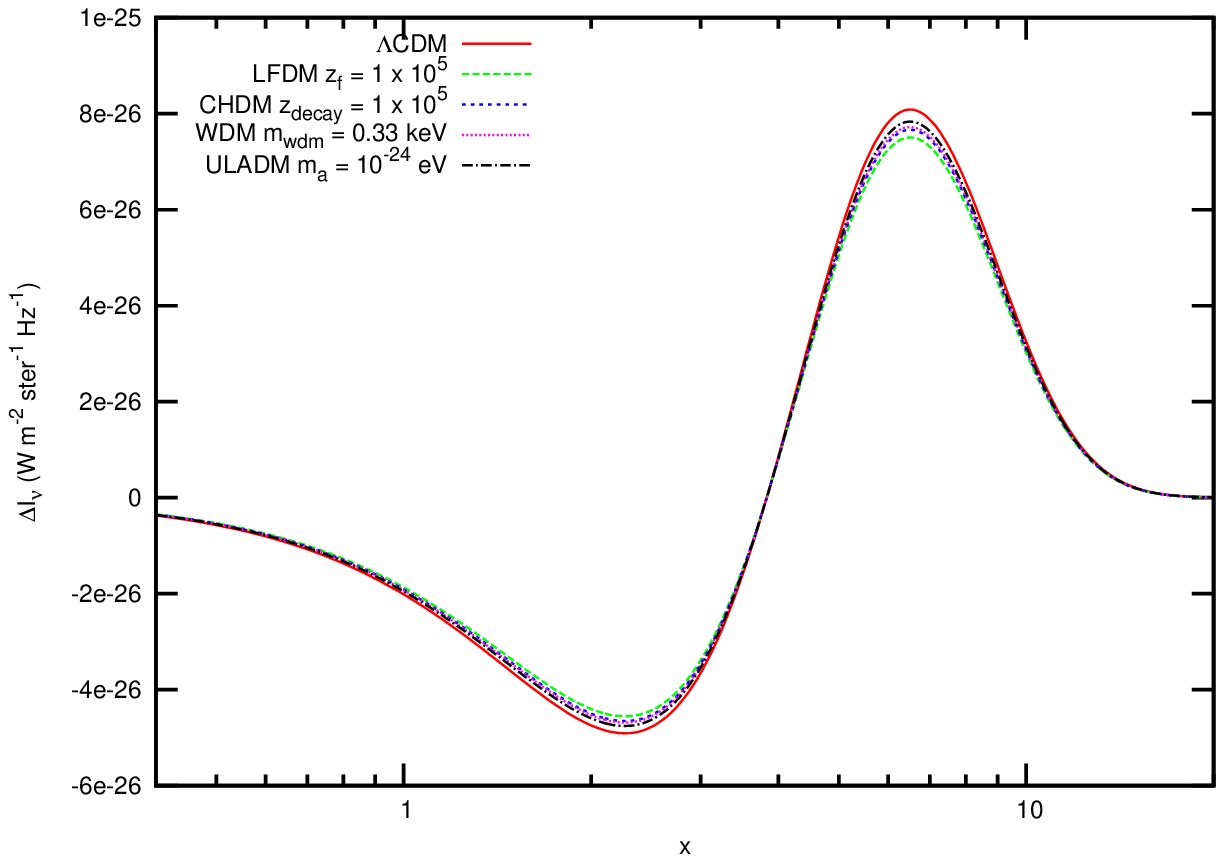}
  \label{fig:1b}
\end{subfigure}
\caption{Comparison of $\Delta\Theta_{1\rm{D}}$(Left) and $y$-parameter(Right) for four different dark matter candidates having  small scale power suppressed at same value of $k \sim 0.3 \rm{h Mpc^{-1}}$. }
\label{fig5}
\end{figure}

Can the $y$-distortion be used to distinguish between different dark matter models? We address this question by comparing models where the power is cut at nearly the same scale. In Figure~\ref{fig5}, we show the evolution of $\Delta\Theta_{1\rm{D}}$ and the  spectral shape of the distorted spectrum for models for which the power has been cut at $k \simeq 0.3 h \, \rm Mpc^{-1}$. \textcolor{black}{Figure~\ref{fig5} also shows the distorted spectrum from $y$-distortion for these cases; the spectra have the same shape (Eq.~(\ref{eq1.8})) but differ owing to slightly different values of the $y$-parameter. A more detailed analysis based on understanding degeneracies between parameters of different models would be needed to quantify the results shown in Figure~\ref{fig5}. 
We also note that this effect can be masked by several late-time phenomena like the Epoch of Reionization, thermal Sunyaev-Zel'dovich(tSZ) effects in the galaxy groups and clusters that give rise to the spectral shape that arises from  $y$-type distortion with orders of magnitude higher $y$-parameters.}

Many of the models we consider are already tightly constrained by cosmological
observations. From galaxy clustering and CMB anisotropy data  and observed neutral hydrogen (HI)  abundance at high redshifts, the LFDM models require $z_f \ga 10^5$ \cite{Sarkar:2014bca,Sarkar:2015dib}. Lyman-$\alpha$ data puts even stronger constraints on such
models \cite{Sarkar:2014bca}.  Recent studies of ULA models suggest that
$m_a \gtrsim 10^{-22}$~eV is consistent with current observational data 
 \cite{Hui:2016ltb}. Other studies based on the abundance of HI at high redshift constrain the axion mass $m_a \gtrsim 10^{-23} \, \rm eV$ \cite{Sarkar:2015dib}. These constraints suggest that some of the models shown in Table~\ref{tab1} are ruled out and in particular the ones that show largest deviations from the $\Lambda$CDM model. However, there is significant variation in the predictions of 
different models, and we might expect to see up to 5\% deviation (e.g. 
$z_f \simeq 10^{5}$)  within 
the framework of constraints that arise from galaxy clustering \cite{Agarwal:2014qca} and CMB 
anisotropies \cite{Hlozek:2016lzm}.

The charged decaying particle model offers a more complicated scenario of spectral distortion of the  CMB. As noted in section~\ref{sec:chdm}, the relativistic electron released as the decay product rapidly thermalizes its energy with the thermal plasma. Eq.~\ref{eq:chdistor} gives the fractional increase in the photon energy density due to this process. This constrains the mass difference between the two heavy particles $\Delta M$ to be tiny. This scenario therefore presents two different ways of distinguishing this model from the $\Lambda$CDM model. 
For instance, for $z_{\rm decay} \simeq 10^5$, the decay will cause $i$-type spectral
distortion in the CMB (whose amplitude will depend on $\Delta M$) while Silk
damping causes a difference in  $y$-distortion. In principle, given
their distinct spectral signatures,  these two phases of distortions are distinguishable from each other\cite{Chluba:2013wsa, Chluba:2013pya}. 

The upcoming experiment Primordial Inflation Explorer (PIXIE) \cite{Kogut:2011xw} is likely to improve the FIRAS bounds on CMB spectral distortion by
many orders of magnitude. Its projected sensitivity corresponds to: $y \simeq 2 \times 10^{-9}$ and $\mu \simeq 10^{-8}$. However, unlike the $\mu$- and $i$-distortion, the $y$-distortion created during the pre-recombination era can be masked by CMB distortion in the post-recombination era.  For 
instance, the epoch of reionization is likely  to produce global  CMB distortion  corresponding to  $y \simeq \hbox{a  few} \times 10^{-7}$ 
\textcolor{black}{, though the most dominating source of post-recombination $y$-distortion is the tSZ effect in galaxy groups and clusters corresponding
to  $y \simeq 2 \times 10^{-6}$\cite{Hill:2015tqa}. These sources would constitute a strong foreground to the pre-recombination  $y$-distortion 
\cite{Kogut:2011xw}. In principle, it might be possible to distinguish 
the pre-recombination $y$-distortion from reionization signal using spatial information (e.g. \cite{Pajer:2013oca, Chluba:2012gq}) but it would be a challenge.}

\section{Conclusion} \label{conclusion}

The large-scale behaviour of dark matter has been well established by a range of cosmological observables such as CMB anisotropies and clustering of galaxies. However, there remain issues with this model at small scales, and the nature of the dominant fraction of dark matter in the universe remains a mystery. 

The evolution of spectral distortion in the pre-recombination era allows us to study scales in the range $10^4 \, {\rm Mpc^{-1}} < k < 0.3 \, \rm Mpc^{-1}$ as density perturbations at these scale decay and leave observable signatures on CMB spectrum. These perturbations are in the linear regime of their growth in the pre-recombination and therefore can be theoretically modelled very accurately. 

We consider four alternative dark matter models which give significant deviations of small-scale power as compared to the $\Lambda$CDM model and study the spectral distortion of CMB owing to Silk damping for these models. These models are motivated by different aspects of physics in the early universe:  phase transition (LFDM), free-streaming of massive particles (WDM), the decay of massive charged particles(CHDM), and dynamics of a scalar field with nonzero effective mass (ULA). 

We show that main impact of the models we consider is to alter the late time spectral distortion history by lowering the $y$-parameter by a few percent for an acceptable range of parameters for these models (Table~\ref{tab1}).

In this work, we only consider models that leave invariant the matter-radiation equality. This excludes models such as decaying dark matter particles that create  relativistic decay products thereby delaying the matter-radiation equality epoch  (e.g. \cite{1998ApJS..114...37B,2004PhRvL..93g1302I}).  We also excluded   models that are based on extra relativistic degrees of freedom (for details
see  \cite{Ade:2015xua} and references therein) or are based on  a change in initial conditions (e.g.  \cite{Chluba:2012we,Clesse:2014pna}).

Our analysis suggests that all class of models that give suppression of power as compared to the $\Lambda$CDM model should result in late time spectral distortion. In section~\ref{cosmology} we present general arguments which show that the change in spectral distortion is dominated by the evolution of potential after the scale enters the horizon. This effect scales as the ratio of energy densities of dark matter and radiation $\rho_{\rm dm}/\rho_r$ and is negligible at early times. This might also allow one to distinguish the impact of modifying the initial matter power spectrum on CMB spectral distortion from a change in the dark matter model that affects the matter power at small scales, as the former would cause all the three forms of CMB distortions while the latter would not.

Future experiments such as PIXIE and CMB-S4 have the potential to establish the nature of dark matter (e.g. \cite{Hlozek:2016lzm}) by unprecedented improvement in mapping the  CMB spectral and spatial structures. Our work is one step in that direction and points out the challenges involved in such an endeavour.

\section*{Acknowledgements}

We thank the anonymous referee for sending useful comments and suggestions that can make the paper better. The authors would like to acknowledge Rishi Khatri, Jens Chluba, David J E Marsh, Ayuki Kamada, Sunny Vagnozzi and Farinaldo Queiroz for their useful comments and suggestions.

\bibliographystyle{JHEP}
\bibliography{sd_notes_final}{}
\end{document}